\documentclass[journal]{IEEEtran}

\ifCLASSINFOpdf
\else
\fi

\usepackage{booktabs}
\usepackage{amsmath,amsfonts}
\usepackage{amsthm}
\usepackage{algorithmic}
\usepackage{verbatim}
\usepackage{graphicx}
\usepackage{cite}
\usepackage[ruled,linesnumbered]{algorithm2e}
\usepackage{cancel}
\usepackage{indentfirst}
\usepackage{epstopdf}
\usepackage{caption}
\usepackage{color}
\usepackage{subfigure}
\usepackage{cite}
\newtheorem{remark}{Remark} 
\usepackage[colorlinks,citecolor=blue,linkcolor=blue,urlcolor=blue]{hyperref}
\begin{document}
	
	\title{Fluid Antenna Enabled Direction-of-Arrival Estimation Under Time-Constrained Mobility}
	\author{He Xu, Tuo Wu, Ye Tian, Kangda Zhi, Wei Liu, Baiyang Liu,  Hing Cheung So, \emph{Fellow, IEEE}, \\  Naofal Al-Dhahir, \emph{Fellow, IEEE},   Kin-Fai Tong,  \emph{Fellow, IEEE},	\\ Chan-Byoung Chae, \emph{Fellow, IEEE},   and  Kai-Kit Wong, \emph{Fellow}, \emph{IEEE}  
		
		\thanks{(\textit{Corresponding author: Tuo Wu.})}
		\thanks{ 
			H. Xu is with the School of Cyber Science and Engineering, Ningbo University of Technology, Ningbo 315211, China (E-mail: $\rm xuhebest@sina.com$).
			Y. Tian is with the Faculty of Electrical Engineering and Computer Science, Ningbo University, Ningbo 315211, China (E-mail: $\rm tianye1@nbu.edu.cn$).
			T. Wu is with the School of Electrical and Electronic Engineering, Nanyang Technological University 639798, Singapore (E-mail: $\rm tuo.wu@qmul.ac.uk$).
			K. Zhi,  and G. Caire are with Communications and Information Theory Group (CommIT), Technische Universitat Berlin, 10587 Berlin, Germany (E-mail: $\rm \{k.zhi, caire\}@tu$-$\rm berlin.de$).
			W. Liu is with the Department of Electrical and Electronic Engineering, Hong Kong Polytechnic University, Kowloon, Hong Kong (E-mail: $\rm wei2.liu@polyu.edu.hk$). 
			B. Liu and  K. F. Tong are with the School of Science and Technology, Hong Kong Metropolitan University, Hong Kong SAR, China. (E-mail: $\rm \{byliu,ktong\}@hkmu.edu.hk$). 
			H. C. So is  with the Department of Electrical Engineering City University of Hong Kong, Hong Kong, China. (E-mail: $\rm hcso@ee.cityu.edu.hk$).
			Naofal Al-Dhahir is with the Department of Electrical and Computer Engineering, The University of Texas at Dallas, Richardson, TX 75080 USA (E-mail: $\rm aldhahir@utdallas.edu$).
			C.-B. Chae is with the School of Integrated Technology, Yonsei University, Seoul 03722 Korea. (E-mail: $\rm cbchae@yonsei.ac.kr$).
			K.-K. Wong is with the Department of Electronic and Electrical Engineering, University College London, WC1E 6BT London, U.K., and also with the Yonsei Frontier Laboratory and the School of Integrated Technology, Yonsei University, Seoul 03722, South Korea (E-mail:$\rm  kai$-$\rm kit.wong@ucl.ac.uk$).
	}} 
	\markboth{IEEE TRANSACTIONS ON SIGNAL PROCESSING,~Vol.~XX, No.~XX, XX~2025}
	{Shell \MakeLowercase{\textit{et al.}}: A Sample Article Using IEEEtran.cls for IEEE Journals}
	\maketitle

	\begin{abstract}
		Fluid antenna (FA) technology has emerged as a promising approach in wireless communications due to its capability of providing increased degrees of freedom (DoFs) and exceptional design flexibility. This paper addresses the challenge of direction-of-arrival (DOA) estimation for aligned received signals (ARS) and non-aligned received signals (NARS) by designing two specialized uniform FA structures under time-constrained mobility. For ARS scenarios, we propose a fully movable antenna configuration that maximizes the virtual array aperture, whereas for NARS scenarios, we design a structure incorporating a fixed reference antenna to reliably extract phase information from the signal covariance. To overcome the limitations of large virtual arrays and limited sample data inherent in time-varying channels (TVC), we introduce two novel DOA estimation methods: TMRLS-MUSIC for ARS, combining Toeplitz matrix reconstruction (TMR) with linear shrinkage (LS) estimation, and TMR-MUSIC for NARS, utilizing sub-covariance matrices to construct virtual array responses. Both methods employ Nyström approximation to significantly reduce computational complexity while maintaining estimation accuracy. Theoretical analyses and extensive simulation results demonstrate that the proposed methods achieve underdetermined DOA estimation using minimal FA elements, outperform conventional methods in estimation accuracy, and substantially reduce computational complexity.
	\end{abstract}
	
	
	\IEEEpeerreviewmaketitle
	
	\section{Introduction}

	\IEEEPARstart{D}{irection}-of-arrival (DOA)  estimation plays a crucial role in various military and civilian applications. In wireless communications, accurate DOA estimation is essential for acquiring channel state information (CSI) and performing effective downlink beamforming \cite{ref1,rr1,ref2}, while also enhancing target detection and tracking capabilities in radar and sonar systems \cite{ref3}.
	
	To date, numerous excellent DOA estimation methods have been proposed under different frameworks. These include subspace-based approaches such as multiple signal classification (MUSIC) \cite{ref4} , principal-singular-vector utilization for modal analysis (PUMA) \cite{add6} and rank-reduction (RARE) based estimators \cite{ref5}, sparse signal reconstruction methods like sparse Bayesian learning (SBL) \cite{ref6} and unitary approximate message passing (UAMP) \cite{ref7}, as well as deep learning-based solutions utilizing convolutional neural networks (CNN) \cite{ref8,rr7,ref9}. However, it is important to note that all these methods are established on fixed-position antennas (FPAs) with element spacing no smaller than half the carrier wavelength. Such FPAs inherently suffer from two major limitations: first, the small antenna spacing leads to significant mutual coupling effects, which severely degrade both communication and parameter estimation performance \cite{ref10}; second, the static steering vector of FPA arrays corresponds to a fixed number of degrees-of-freedom (DoFs), which not only weakens secure beamforming gain but also restricts the implementation of super-resolution estimation and the ability to detect a large number of sources in practice. While massive multiple-input multiple-output (MIMO) antenna arrays can address these issues, their higher hardware costs are neither economically viable nor sustainable.
	
	As an alternative solution, fluid antenna systems (FAS), also known as movable antenna systems, have emerged as a promising approach in recent years, offering the potential to overcome the limitations of FPA systems \cite{ref11,refnew11,refnew12,ref12,ref13,ref14}. In an FA system, the position of the radiating element can be dynamically reconfigured across a spatial aperture, enabling flexible and adaptive control over the antenna’s spatial behavior. This reconfigurability can be achieved through various means—such as electronically switchable pixel arrays, metasurfaces, or other tunable structures—without necessarily requiring any physical displacement or fluidic materials. By allowing the effective radiation point to change in response to the environment or communication needs, FA systems unlock additional spatial DoFs, offering new opportunities for enhancing performance in next-generation wireless systems.  Motivated by these advantages, several FA-enabled techniques and schemes have been investigated from various perspectives, including FA multiple access (FAMA) \cite{add13,add14,add15}, channel estimation \cite{add16,add17,add18,aadd18}, beamforming design \cite{add19,add20}, and integrated sensing and communications (ISAC) \cite{add21,add22}.

	While these studies demonstrated FA's versatility across various communication applications, its potential for enhancing direction-finding capabilities remains largely unexplored. FA offers several key advantages for DOA estimation. First, the dynamic movement of FAs enables the construction of larger virtual arrays within limited mobility time, significantly increasing spatial DoFs. This expanded virtual array not only enhances estimation accuracy but also enables underdetermined DOA estimation, where the number of estimable sources can exceed the number of physical antennas. Second, the flexible positioning capability of FAs allows for adaptive array configurations that can be optimized for different scenarios, providing improved spatial resolution and better discrimination between closely spaced sources.
	
	Despite these significant benefits, the mobility of FA also introduces unique challenges in real-world applications. To understand these challenges, it is essential to clarify the temporal framework of DOA estimation employing FAs. The entire DOA estimation process operates within a larger time period during which the target directions remain constant. This total observation period is divided into multiple consecutive channel coherence time blocks, within each of which the complex channel gains are assumed to remain approximately constant. However, each individual coherence time block is relatively short, creating the fundamental time-constrained mobility problem. Specifically, FAS faces several critical timing constraints: all FA movements within a single channel coherence time must complete before the channel gains change; each position measurement must capture a statistically reliable sample under stable channel conditions; and the total observation period is limited by how long the target maintains its angular direction \cite{add23,add24,ref16}. These inherent limitations impose strict constraints on system design, where the finite mobility speed of FA elements restricts the achievable positions within each coherence time block, while the time-varying nature of wireless channels limits the duration available for spatial sampling.  
	
	Beyond these timing constraints, another critical challenge arises from the nature of received signals in FAS. The key distinction lies in whether the transmitted signal content remains identical or varies across different FA positions within the same time block. This fundamental difference determines the processing strategy and directly impacts the achievable performance of DOA estimators.
	
	In aligned received signal (ARS) scenarios, e.g.,  when known pilot sequences are transmitted, the signal content remains identical across all FA positions within a time block, enabling direct combination of received samples from different FA positions to form a coherent virtual array. This coherent virtual array is an extended array formed by FA movements across different positions, where signals from different locations maintain fixed phase relationships and can be processed coherently like a traditional physical array \cite{reff161}. Mathematically, for two FA positions within a time block, the received signals can be expressed as $x_1 = a_1 s + n_1$ and $x_2 = a_2 s + n_2$, where the same signal $s$ is observed at both positions with different spatial responses $a_1$ and $a_2$. This enables direct stacking: $[x_1, x_2]^T = [a_1, a_2]^T s + [n_1, n_2]^T$, forming a virtual array with extended response vector $[a_1, a_2]^T$.
	
	In contrast, non-aligned received signal (NARS) scenarios present a more challenging situation. When unknown data symbols are transmitted, the signal content varies independently across different FA positions within the same time block, preventing direct signal combination. Considering the above-mentioned two-position example, NARS case involves different signals: $x_1 = a_1 s_1 + n_1$ and $x_2 = a_2 s_2 + n_2$, where $s_1 \neq s_2$. Direct stacking becomes impossible since there is no common signal to factor out. However, this issue can be overcome through covariance-based approaches that extract spatial information from signal statistics rather than from the signal samples themselves. The key insight is that while the signals themselves differ, their cross-covariance $\mathbb{E}\{x_1 x_2^*\} = \mathbb{E}\{|s|^2\} a_1 a_2^* e^{-j(d_1-d_2)\frac{2\pi}{\lambda}\sin\theta}$ still preserves the crucial phase information $e^{-j(d_1-d_2)\frac{2\pi}{\lambda}\sin\theta}$ that depends only on the position difference $(d_1-d_2)$ and the DOA $\theta$ \cite{reff162,reff163,reff164}. By collecting such covariance elements across different position differences, we can construct a virtual array response that enables DOA estimation without requiring signal alignment.
	
	Given these fundamentally different signal characteristics, the antenna movement patterns must be carefully designed for each scenario. For ARS, the movement strategy should generate consecutive spatial sampling points to maximize the virtual array aperture within limited mobility time. This consecutive spatial sampling is crucial because direct signal stacking is allowed, where the goal is to construct a standard virtual uniform linear array (ULA) to obtain the well-known Vandermonde structure so that classical array processing
	algorithms \cite{reff165,reff166} can be applied.  For NARS, however, a more sophisticated approach is needed, typically involving a fixed reference antenna combined with movable elements. The fixed reference antenna serves as a stable baseline for covariance computation, ensuring that rich second-order difference lags (i.e., position differences between antenna pairs) can be generated to adequately sample the spatial correlation function. This design consideration is crucial for producing an effective virtual array that can extract the necessary phase information for DOA estimation despite the non-aligned nature of the signals.

	In this paper, to address the aforementioned challenges in FA mobilities and signal characteristics, two FAS-enabled DOA estimation methods designed for both ARS  and  NARS scenarios are proposed, under the limited mobility time conditions. The main contributions are:
	\begin{itemize}
		\item \textit{\textbf{Optimized FA movement patterns:}}
		\begin{itemize}
			\item For ARS scenarios: we design a uniform FA array structure where all elements are movable, creating a virtual ULA with consecutive spatial sampling points. This maximizes the virtual array aperture within limited mobility time, directly enhancing spatial resolution while allowing direct application of conventional DOA estimation algorithms.
			\item For NARS scenarios: we propose a specialized configuration featuring one fixed reference antenna combined with FAs. This pattern is specifically devised to generate rich second-order difference lags between antenna positions, enabling effective sampling of the spatial correlation function even when signals cannot be directly combined.
		\end{itemize}
		\item \textit{\textbf{Enhanced covariance estimation techniques:}}
		\begin{itemize}
			\item For ARS scenarios: we develop the TMRLS-MUSIC algorithm combining Toeplitz matrix reconstruction (TMR) with linear shrinkage (LS) estimation to address the ``high-dimensional, comparable-sample" problem created by the large virtual array. This approach adaptively balances between structured and unstructured covariance estimates to optimize performance across varying signal-to-noise ratio (SNR) conditions.
			\item For NARS scenarios: we introduce the TMR-MUSIC algorithm that constructs a virtual array response from sub-covariance matrices, enabling DOA estimation despite non-aligned signals. Our method extracts phase information from the covariance structure rather than the signals themselves, making it robust to variations in transmitted signals.
		\end{itemize}
		\item \textit{\textbf{Comprehensive performance validation:}} Performance studies involving  estimation accuracy, maximum number of estimable users or targets (U/Ts), computational complexity, and ability to localize U/Ts in different transmission scenarios are provided. Extensive experimental results under various   configurations demonstrate that our proposed solutions can achieve underdetermined DOA estimation and yield improved performance with reduced complexity, facilitating their practical applicability.
	\end{itemize}
	
	The rest of this paper is organized as follows. Section \ref{signal model} describes the signal model for FAS-enabled DOA estimation in time-varying block fading model. Section \ref{DOA} introduces the designed FA array structures, as well as corresponding DOA estimation methods for both scenarios of ARS and NARS in detail. Performance analysis are given in Section~\ref{performance}. Simulation results are provided in Section~\ref{Simulation}. Finally, the conclusions are drawn in Section~\ref{conclusions}.
	
	{\emph{Notations:} ${\bf{A}}$, $\mathcal{A}$, ${\bf{a}}$ and $a$ stand for a matrix, a set, a vector and a scalar, respectively. ${\left\|  \cdot  \right\|_F}$ stands for the Frobenius norm, and $\mathbb{E}\{  \cdot \} $ the statistical expectation. ${\mathop{\rm diag}\nolimits} \left(  \cdot  \right)$, ${\mathop{\rm blkdiag}\nolimits} \left(  \cdot  \right)$,
		${\left(  \cdot  \right)^T}$ and ${\left(  \cdot  \right)^H}$ denote the diagonalization, block diagonalization, transpose and conjugate transpose operators, respectively. ${\mathop{\rm Tr}\nolimits} \left(  \cdot  \right)$ represents the trace of a matrix, ${\bf{I}}_M$ is the $M\times M$ identity matrix, ${\bf{J}}^m$ the shift matrix whose elements on the $m$-th super diagonal are one and zeros otherwise, ${{\bf{\Pi }}_M}$ the $M\times M$ exchange matrix with all ones on its anti-diagonal positions and zeros elsewhere, 
		and finally, $\mathcal{CN}\left( {{\bf{a}},{\bf{A}}} \right)$ denotes the complex Gaussian distribution with mean vector $\bf{a}$ and covariance matrix $\bf{A}$.

		\section{Signal Model}\label{signal model}
		Consider a far-field narrowband millimeter wave (mmWave) system consisting of $K$ single-antenna U/Ts transmitting signals to a receiver equipped with an $M$-element uniform linear  FA array. Due to the inherent limited scattering characteristics of mmWave propagation environments, we model each U/T as being associated with $L$ dominant scatterers, with each scatterer contributing exactly one unique propagation path to the channel. Throughout the entire DOA estimation process, it is assumed that the  DOA  associated with each propagation path remains invariant, which is reasonable   for the considered limited mobility time. For practical channel modeling, we adopt a time-varying block fading model \cite{ref17,ref18}, wherein the complex channel gain for each path remains constant within each short time block, but may vary independently from one block to another, capturing the dynamic nature of wireless channels. Based on these assumptions, the uplink channel vector ${\bf h}_{k,n,g}$ for the $k$-th U/T in the $n$-th time block, after the $g$-th movement of the FA array, can be expressed as
		\begin{equation}\label{1}
			{{\bf{h}}_{k,n,g}} = \sum_{l = 1}^L {{\alpha _{k,n,l}}} {{\bf{a}}_g}\left( {{\theta _{k,l}}} \right) = {{\bf{A}}_{k,g}}{{\boldsymbol{\alpha }}_{k,n}},
		\end{equation}
		where $k\in [1,K], n\in [1,N]$, $g\in[0,G]$, ${\alpha_{k,n,l}} \sim \mathcal{CN}\left( {0,{\sigma_\alpha^2}} \right)$ denotes the complex path gain associated with the $l$-th scattering path of the $k$-th signal in the $n$-th time block, ${\boldsymbol{\alpha}}_{k,n} = {\left[ {{\alpha_{k,n,1}}, \ldots ,{\alpha_{k,n,L}}} \right]^T}$ represents the vector of all path gains for the $k$-th U/T, and ${{\bf{A}}_{k,g}} = \left[ {{\bf{a}}_g\left( {{\theta_{k,1}}} \right), \ldots ,{\bf{a}}_g\left( {{\theta_{k,L}}} \right)} \right]$ is the array manifold matrix whose $l$-th column ${\bf{a}}_g\left( {{\theta_{k,l}}} \right)$ is given by 
		\begin{equation}\label{2}
			{{\bf{a}}_g}\left( {{\theta _{k,l}}} \right) = {\left[ {{e^{ - j\left( {{d_{1,g}} - {d_0}} \right){\varphi _{k,l}}}}, \ldots ,{e^{ - j\left( {{d_{M,g}} - {d_0}} \right){\varphi _{k,l}}}}} \right]^T},
		\end{equation}
		with ${\varphi _{k,l}} = \frac{{2\pi }}{\lambda }\sin \left( {{\theta _{k,l}}} \right)$, where $\lambda$ denotes the carrier wavelength and $\theta _{k,l}$ represents the physical DOA of the $l$-th path from the $k$-th U/T. Moreover, $d_{m,g}\left( {1 \le m \le M} \right)$ indicates the coordinate of the $m$-th FA element along the $x$-axis after the $g$-th movement, and $d_0$ is defined as the coordinate reference point.
		\begin{figure}[t]
			\centering 
			\includegraphics[width=3.6in]{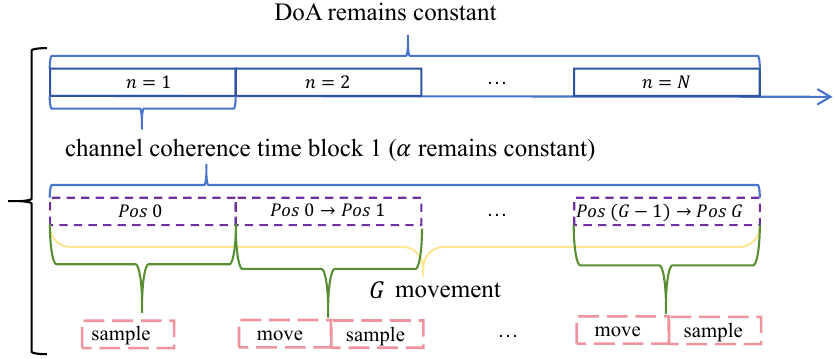} 
			\caption{Illustration of Time-Constrained FA Mobility for DOA Estimation.}\label{Mobility}
		\end{figure}
		
		To better illustrate this temporal structure and mobility constraints, Fig.~\ref{Mobility} presents the time-constrained FA mobility framework, where the hierarchical temporal organization of DOA estimation in FA system is demonstrated: the entire process operates within a total observation period (during which target directions remain constant), which is subdivided into $N$ consecutive channel coherence time blocks. Within each block $n$, the FA array must complete $G$ sequential movements from position 0 to position $G$, generating $G+1$ signal snapshots per block, while ensuring that channel gains remain approximately constant. This temporal hierarchy directly illustrates the time-constrained mobility problem, where the relatively short duration of each coherence time block fundamentally limits the spatial DoFs that can be achieved through FA movements.

		Suppose that $K$ equal-power transmitted signals by U/Ts are ${{\bf{s}}_{n,g}} = {\left[ {{s_{1,n,g}}, \ldots ,{s_{K,n,g}}} \right]^T}$. Considering the practical challenges in FA systems with limited mobility time, we distinguish between two critical scenarios that significantly affect DOA estimation performance:
		
		1)  ARS: when the transmitted signals are pilots, their elements are known and can be aligned at the receiver (i.e., $s_{k,n,g_1}=s_{k,n,g_2}$, $g_1\neq g_2$), enabling traditional array processing techniques with extended virtual arrays;
		
		2)  NARS: when the transmitted signals are data transmission (DT) ones, they are unknown and inherently non-aligned across different FA positions (i.e., $s_{k,n,g_1}\neq s_{k,n,g_2}$, $g_1\neq g_2$), necessitating novel processing approaches. 
		
		This distinction, often overlooked in existing literature, is crucial since most practical FA systems could operate in both scenarios, particularly when monitoring non-cooperative targets or during regular communication sessions. We assume the transmitted signals are independent of complex path gains. Subsequently, the array output corresponding to the $n$-th time block after the $g$-th movement can be expressed as
		\begin{multline}\label{3}
			{{\bf{x}}_{n,g}} = \left[ {{{\bf{h}}_{1,n,g}}, \ldots ,{{\bf{h}}_{K,n,g}}} \right]{{\bf{s}}_{n,g}} + {{\bf{e}}_{n,g}} \\
			= {{\bf{A}}_g}{{\bf{D}}_{n}}{{\bf{s}}_{n,g}} + {{\bf{e}}_{n,g}}={{\bf{A}}_g}{{{\bf{s'}}}_{n,g}}+ {{\bf{e}}_{n,g}},\quad\quad\quad\quad
		\end{multline}
		where ${{{\bf{s'}}}_{n,g}} = {{\bf{D}}_n}{{\bf{s}}_{n,g}} \in {\mathbb{C}^{LK \times 1}}$ represents the effective transmitted signal vector modulated by the channel gains, whose elements still obey complex Gaussian distribution due to the features of path gains, and ${{\bf{e}}_{n,g}} \sim \mathcal{CN}\left( {0,\sigma _e^2{{\bf{I}}_M}} \right)$ is the additive white Gaussian noise vector.   ${{\bf{D}}_{n}} = {\mathop{\rm blkdiag}\nolimits} \left( {\left[ {{\boldsymbol \alpha _{1,n}}, \ldots ,{\boldsymbol \alpha _{K,n}}} \right]} \right)$ is block diagonal, containing the complex path gains ${{\boldsymbol \alpha }_{k,n}}$ for all users in the $n$-th time block.   ${{\bf{A}}_g} = [{{\bf{A}}_{1,g}}, \ldots ,{{\bf{A}}_{K,g}}]$ serves as the overall array manifold matrix, constructed by concatenating the individual user manifold matrices ${{\bf{A}}_{k,g}} = [{{\bf{a}}_g}({\theta _{k,1}}), \ldots ,{{\bf{a}}_g}({\theta _{k,L}})]$. Each ${{\bf{A}}_{k,g}}$ encapsulates the steering vectors ${{\bf{a}}_g}({\theta _{k,l}})$ corresponding to the $L$ propagation paths of the $k$-th U/T after the $g$-th movement.

		We further assume that the FA's positioning system is capable of implementing $G$ movements within a short time block. This allows us to express the total output related to the $n$-th time block as
		\begin{equation}\label{4}
			{{\bf{x}}_n} = \left[ \begin{array}{l}
				{{\bf{x}}_{n,0}}\\
				\quad \vdots \\
				{{\bf{x}}_{n,G}}
			\end{array} \right] = \left[ \begin{array}{l}
				{{\bf{A}}_0}\\
				\quad\quad \ddots \\
				\quad\quad\quad\quad{{\bf{A}}_G}
			\end{array} \right]\left[ \begin{array}{l}
				{{{\bf{s'}}}_{n,0}}\\
				\quad \vdots \\
				{{{\bf{s'}}}_{n,G}}
			\end{array} \right] + {{\bf{e}}_n},
		\end{equation}
		where ${{\bf{e}}_n} = {\left[ {{\bf{e}}_{n,0}^T, \ldots ,{\bf{e}}_{n,G}^T} \right]^T}$.
		
		Specifically, for the ARS scenario where the transmitted signals are identical across movements within a block (i.e., ${{\bf{s}}_{n,g}}$ is constant for $g=0,\dots,G$), the effective signal vector also remains constant: ${{{\bf{s'}}}_n} = {{{\bf{s'}}}_{n,0}} = \cdots = {{{\bf{s'}}}_{n,G}}$. Consequently, Eq. (\ref{4}) simplifies to
		\begin{equation}\label{5}
			{{\bf{x}}_n} = {\bf{A}}{{{\bf{s'}}}_n} + {{\bf{e}}_n},
		\end{equation}
		where the matrix ${\bf{A}} = {[{\bf{A}}_0^T, \ldots ,{\bf{A}}_G^T]^T}$ represents the extended virtual array manifold matrix formed by vertically stacking the manifold matrices from all $G$ movements \footnote{It is important to note that this simplified expression (\ref{5}) is valid only for ARS. The NARS scenario, where the effective signal vectors ${\bf s'}_{n,g}$ vary with $g$ and thus cannot be factored out, requires a different approach based on covariance matrices, which will be detailed in the following Section~\ref{sec:NARS_model}.  }.
		
		By collecting the outputs ${\bf x}_n$ from Eq. (\ref{5}) across $N$ consecutive time blocks, we obtain the overall data:
		\begin{equation}\label{6}
			{\bf{X}} = \left[ {{{\bf{x}}_1}, \ldots ,{{\bf{x}}_N}} \right] = \left[ {\begin{array}{*{20}{l}}
					{{{\bf{A}}_0}}\\
					{\quad \quad  \ddots }\\
					{\quad \quad \quad \quad {{\bf{A}}_G}}
			\end{array}} \right]{\bf{S}} + {\bf{E}}
		\end{equation}
		where ${\bf{S}} = \left[ {{{{\bf{s'}}}_1}, \ldots ,{{{\bf{s'}}}_N}} \right]$ and ${\bf{E}} = \left[ {{{\bf{e}}_1}, \ldots ,{{\bf{e}}_N}} \right]$.
		
		In the subsequent sections, we will elaborate on how to utilize  ${\bf{X}}$ to perform DOA estimation under both ARS and NARS scenarios.

		\section{DOA Estimation}\label{DOA}
		In this section, we first describe the design principle of uniform FA array structures for ARS and NARS, respectively, and then exploit them for DOA estimation through the TMR and Nystr\"{o}m approximation based subspace techniques.
		\subsection{FA Array Design and DOA Estimation for ARS}\label{sec:ARS_model}
		
		For the ARS scenario, our goal is to design an FA movement strategy that maximizes the consecutive   DoFs  or lags achievable within the limited mobility time constraint (i.e., $G$ movements). We propose the uniform FA array structure and movement pattern illustrated in Fig. \ref{Fig.antennas_align}. Here, $d \leq \lambda/2$ denotes the basic movement step size. Without loss of generality, we set the coordinate reference point $d_0 = 0$. The initial positions (at $g=0$) of the $M$ FA elements are set as ${d_{m,0}} = (m-1)(G+1)d$, ($m=1, \dots, M$). This specific initialization, combined with the movement pattern described below, ensures the generation of a large virtual ULA.

		\emph{Proposition 1:} Let ${\mathcal{L}_P} = \left\{ {{l_p}|{l_p}d \in {\mathcal{C}_P}} \right\}$ denote the first-order consecutive DoFs/lags after $G$ movements  where  ${\mathcal{C}_P}$ is the position set of all elements. For the designed FA array structures for ARS, ${\mathcal{L}_P}$ contains $\bar M=MG+M$ consecutive DoFs/lags in the range $0 \le {l_p} \le \bar M - 1$.
		\begin{proof}
			For the $g$-th movement, the produced unique lags become  ${\mathcal{L}_{Pg}} = \{ {{l_p}|\frac{{{l_p} - g}}{{(G + 1)}} \in \left[ {0,M - 1} \right]} \}$, $g\in [0, G]$. Given the fact that $g$ increases continuously with a step size of 1. After $G$ moves, we   obtain that the $m$-th FA will yield $G+1$ consecutive positions within the range $[m(G+1),m(G+1)+G]d$. Finally, by combining the result of all FAs, we easily know that ${\mathcal{L}_P}$ contains $\bar M$ consecutive DoFs/lags in the range $0 \le {l_p} \le \bar M - 1$.
		\end{proof}
		\begin{figure}[t]
			\centering
			\includegraphics[width=3.4in]{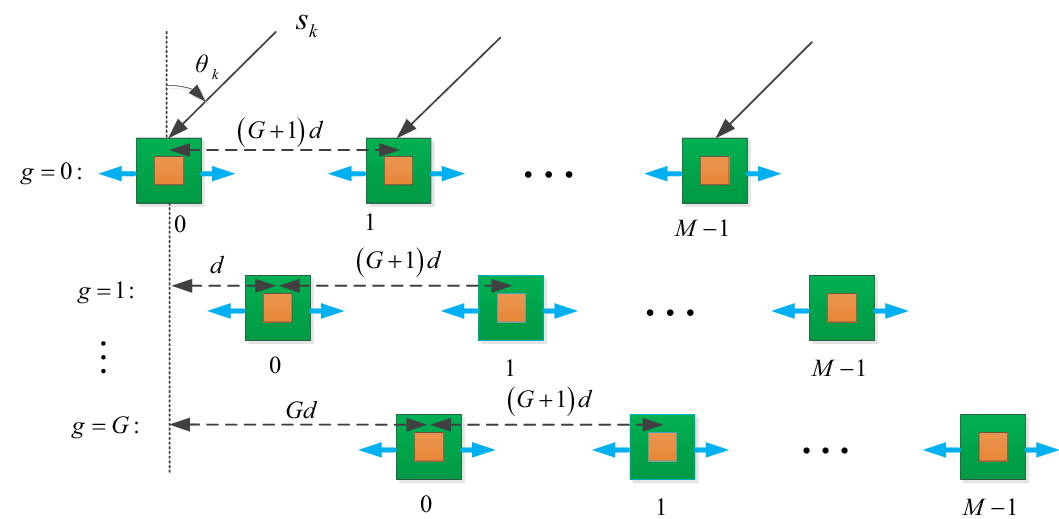}\\
			\caption{Illustration of FA array structures for cases of ARS.}\label{Fig.antennas_align}
		\end{figure}
		
		Proposition 1 reveals that the designed FA movement strategy effectively synthesizes a virtual ULA   comprising $\bar M = M(G+1)$ consecutive elements with spacing $d$. The received data matrix ${\bf X}$ contains measurements corresponding to all these virtual positions. By rearranging the rows of ${\bf X}$ according to the spatial sequence of these virtual elements (from $0$ to $(\bar M - 1)d$), we obtain the $\bar M \times N$ matrix ${\bf Y}$:
		\begin{multline}\label{7}
			{\bf{Y}} = \left[ {{\bf{X}}_{[1]}^T,{\bf{X}}_{[M + 1]}^T, \cdots ,{\bf{X}}_{[GM + 1]}^T, \cdots ,} \right.\\
			\left. {{\bf{X}}_{[M]}^T,{\bf{X}}_{[2M]}^T, \cdots ,{\bf{X}}_{[(G + 1)M]}^T} \right] = {\bf{BS}} + {\bf{\bar E}},
		\end{multline}
		where subscript represents the row index of   $\bf{X}$. In this standard ULA signal model, ${\bf S}$ is the effective signal matrix, ${\bf{\bar E}}$ is the correspondingly rearranged noise matrix ${\bf E}$, and ${\bf B}$ represents the array manifold matrix of the synthesized $\bar M$-element virtual ULA. Consequently, the columns of ${\bf B}$, denoted as ${\bf{b}}(\theta_{k,l})$, adopt the well-known Vandermonde structure characteristic of    ULA, i.e., ${\bf{b}}(\theta) = [1, e^{-j\varphi}, \dots, e^{-j(\bar M-1)\varphi}]^T$ with $\varphi=(2\pi/\lambda)\sin(\theta)$. This equivalence facilitates the direct application of various established ULA-based DOA estimation algorithms to ${\bf Y}$. 
		However, the synthesized virtual array size $\bar M$ can be large, while the number of available snapshots $N$ might be limited due to the FA's mobility constraints and the channel's time-varying nature (TVC). In such high-dimensional, small-sample scenarios, conventional sample covariance matrix (SCM) estimation can be inaccurate and unstable. 
		
		\begin{remark} 
			In the context of FA-enabled DOA estimation, the ``high-dimensional, comparable-sample" problem refers to scenarios where the number of parameters in the covariance matrix ($\mathcal{O}(\bar{M}^2)$) and the available snapshot number ($N$) are of comparable magnitude. Under the general asymptotic theory (GAT) framework \cite{ref20} where $\bar{M}/N \to c > 0$, the conventional sample covariance matrix estimation is no longer unbiased, necessitating specialized techniques to enhance estimation performance.
		\end{remark}
		
		To address this challenge, we employ a robust covariance matrix estimation approach that combines  TMR \cite{ref21} with LS  estimation \cite{ref19}, leading to the TMRLS-MUSIC algorithm proposed herein.   Based on  \eqref{7}, the array covariance matrix of ${\bf{Y}}$ is given by
		\begin{equation}\label{8}
			{{\bf{R}}_Y} = \mathbb{E}\left\{ {{\bf{Y}}{{\bf{Y}}^H}} \right\} = {\bf{B}}{{\bf{R}}_S}{{\bf{B}}^H} + \sigma _e^2{{\bf{I}}_{\bar M}},
		\end{equation}
		where ${{\bf{R}}_S} = \mathbb{E}\left\{ {{\bf{S}}{{\bf{S}}^H}} \right\}\in \mathbb{R}^{KL\times KL}$, and the corresponding sampled covariance matrix (SCM) of ${{\bf{R}}_Y}$ is ${{\bf{\hat R}}_Y} = \frac{1}{N}{\bf{Y}}{{\bf{Y}}^H}$.
		
		According to the GAT framework, the SCM ${{\bf{\hat R}}_Y}$ is no longer an unbiased estimate of ${{\bf{R}}_Y}$ in the case of large-scale arrays and limited sample data. Based on the known Toeplitz structure of ${{\bf{R}}_Y}$ for uniform arrays, we first apply the Toeplitz rectification technique \cite{ref21} to enforce the SCM to have a Toeplitz structure by averaging its entries along the diagonals. This is achieved by using the shift matrix ${\bf{J}}^m$ to extract elements along each diagonal, computing their average, and then reconstructing a matrix with constant values along each diagonal, yielding
		\begin{equation}\label{9}
			{{\bf{R}}_T} = \sum\limits_{m =  - \bar M + 1}^{\bar M - 1} {\frac{1}{{\bar M - \left| m \right|}}} {\mathop{\rm Tr}\nolimits} \left( {{{{\bf{\hat R}}}_Y}{{\bf{J}}^m}} \right){\left( {{{\bf{J}}^T}} \right)^m}.
		\end{equation}
		While the Toeplitz rectification ${{\bf{R}}_T}$ enforces the desired structure, it may also discard some valuable statistical information present in the original SCM ${{\bf{\hat R}}_Y}$. In low SNR regime, the structured ${{\bf{R}}_T}$ can provide robustness, whereas in high SNR regime, the original ${{\bf{\hat R}}_Y}$ might retain more accurate signal characteristics. To leverage the advantages of both estimators and achieve enhanced performance across varying SNRs, we propose a balanced combination through LS. Next, we construct the following LS estimation problem to obtain a more robust estimate of ${{\bf{R}}_Y}$ by finding an optimal combination of ${{\bf{\hat R}}_Y}$ and ${{\bf{R}}_T}$:
		\begin{equation}\label{10}
			\mathop {\min }\limits_\rho  \mathbb{E}\{ {\| {{{{\bf{\tilde R}}}_Y} - {{\bf{R}}_Y}} \|_F^2} \} 
			\quad \textrm{s.t.}\quad{{{\bf{\tilde R}}}_Y} = \left( {1 - \rho } \right){{{\bf{\hat R}}}_Y} + \rho {{\bf{R}}_T},
		\end{equation}
		where $\rho  \in \left[ {0,1} \right]$ represents the LS coefficient. To solve this optimization problem, we need to expand the objective function and find the value of $\rho$ that minimizes the expected mean squared error (MSE).  By setting the first-order derivative of $\mathbb{E}\{ {\| {{{{\bf{\tilde R}}}_Y} - {{\bf{R}}_Y}} \|_F^2} \}$ with respect to $\rho$ to zero, the optimal estimator of LS coefficient $\rho$ is given by
		\begin{equation}\label{11}
			{\rho _O} = \frac{{\mathbb{E}\{ {\mathop{\rm Tr}\nolimits} [{{{\bf{\hat R}}}_Y}({{{\bf{\hat R}}}_Y} - {{\bf{R}}_T})]\}  - \mathbb{E}\{ {\mathop{\rm Tr}\nolimits} [{{\bf{R}}_Y}({{{\bf{\hat R}}}_Y} - {{\bf{R}}_T})]\} }}{{\mathbb{E}\{ {\mathop{\rm Tr}\nolimits} {{({{{\bf{\hat R}}}_Y} - {{\bf{R}}_T})}^2}\} }}.
		\end{equation}
		Notice that ${{{\bf{R}}_Y}}$ is not available in practice, we exploit the following approximations~\cite{ref22},\cite{ref23}:
		\begin{equation}\label{12}
			\mathbb{E}\{ {\mathop{\rm Tr}\nolimits} [{{{\bf{\hat R}}}_Y}({{{\bf{\hat R}}}_Y} - {{\bf{R}}_T})]\}  \approx {\mathop{\rm Tr}\nolimits} [{{{\bf{\hat R}}}_Y}({{{\bf{\hat R}}}_Y} - {{\bf{R}}_T})]
		\end{equation}
		\begin{equation}\label{13}
			\mathbb{E}\{ {\mathop{\rm Tr}\nolimits} [{({{{\bf{\hat R}}}_Y} - {{\bf{R}}_T})^2}]\}  \approx {\mathop{\rm Tr}\nolimits} [{({{{\bf{\hat R}}}_Y} - {{\bf{R}}_T})^2}],
		\end{equation}
		\begin{equation}\label{14}
			\begin{array}{l}
				\mathbb{E}\{ {\mathop{\rm Tr}\nolimits} [{{\bf{R}}_Y}({{{\bf{\hat R}}}_Y} - {{\bf{R}}_T})]\}\\
				= \mathop{\rm Tr} ({\bf{R}}_Y \cdot \mathbb{E}\{{{{\bf{\hat R}}}_Y} - {{\bf{R}}_T})\} \\
				= \mathop{\rm Tr} ({\bf{R}}_Y \cdot \mathbb{E}\{{{{\bf{\hat R}}}_Y}\})- \mathop{\rm Tr} ({\bf{R}}_Y \cdot \mathbb{E}\{{{{\bf{R}}}_T}\})\\
				=\mathop{\rm Tr} ({\bf{R}}_Y \cdot \mathbb{E}\{{{{\bf{\hat R}}}_Y}\})\\
				\quad - {\textstyle \sum_{m =  - \bar M + 1}^{\bar M - 1}} {\frac{1}{{\bar M - \left| m \right|}}} {\mathop{\rm Tr}\nolimits} ( {{{{\bf{\hat R}}}_Y}{{\bf{J}}^m}} ) \mathop{\rm Tr}({\bf{R}}_Y{( {{{\bf{J}}^T}} )^m}).
			\end{array}
		\end{equation}
		Then, given the following asymptotic behaviors in the GAT framework \cite{add25}, \cite{add26}
		\begin{equation}\label{15}
			\mathop{\rm Tr} ({\bf{R}}_Y) =\mathop{\rm Tr} ({\bf{\hat R}}_Y),
		\end{equation}
		\begin{equation}\label{16}
			\mathop{\rm Tr} ({\bf{R}}_Y^2) =\frac{{{{(N - 1)}^2}}}{{(N - 2)(N + 1)}} \{\mathop{\rm Tr} ({\bf{\hat R}}_Y^2)-\frac{{\mathop{\rm Tr}}^2 ({\bf{\hat R}}_Y)}{N-1}\},
		\end{equation}
		it can be derived that
		\begin{equation}\label{17}
			\begin{array}{l}
				\mathbb{E}\{ {\mathop{\rm Tr}\nolimits} [{{\bf{R}}_Y}({{{\bf{\hat R}}}_Y} - {{\bf{R}}_T})]\}\\
				=\mathop{\rm Tr} ({\bf{R}}_Y^2 ) - {\textstyle \sum_{m =  - \bar M + 1}^{\bar M - 1}} {\frac{1}{{\bar M - \left| m \right|}}} {\mathop{\rm Tr}\nolimits} ( {{{{\bf{\hat R}}}_Y}{{\bf{J}}^m}} ) \mathop{\rm Tr}({\bf{\hat R}}_Y{( {{{\bf{J}}^T}} )^m})\\
				\approx \frac{{{{(N - 1)}^2}}}{{(N - 2)(N + 1)}} [\mathop{\rm Tr} ({\bf{\hat R}}_Y^2)-\frac{{\mathop{\rm Tr}}^2 ({\bf{\hat R}}_Y)}{N-1} ] -\mathop{\rm Tr} ({{{\bf{\hat R}}}_Y}{{{\bf{R}}}_T}).
			\end{array}
		\end{equation}
		Substituting \eqref{12}, \eqref{13} and \eqref{17} into \eqref{11}, we have
		\begin{equation}\label{18}
			{\rho _a} = \frac{{(N - 3){\mathop{\rm Tr}\nolimits} ({\bf{\hat R}}_Y^2) + (N - 1){{{\mathop{\rm Tr}\nolimits} }^2}({{{\bf{\hat R}}}_Y})}}{{(N - 2)(N + 1){\mathop{\rm Tr}\nolimits} [{{({{{\bf{\hat R}}}_Y} - {{\bf{R}}_T})}^2}]}},
		\end{equation}
		and finally, the enhanced SCM is calculated by
		\begin{equation}\label{19}
			{{{\bf{\tilde R}}}_Y} = \left( {1 - \hat \rho } \right){{{\bf{\hat R}}}_Y} + \hat \rho {{\bf{R}}_T},
		\end{equation}
		with $\hat \rho  = \min \left( {{\rho _a},1} \right)$.
		\begin{figure}[t]
			\centering
			\includegraphics[width=3.5in]{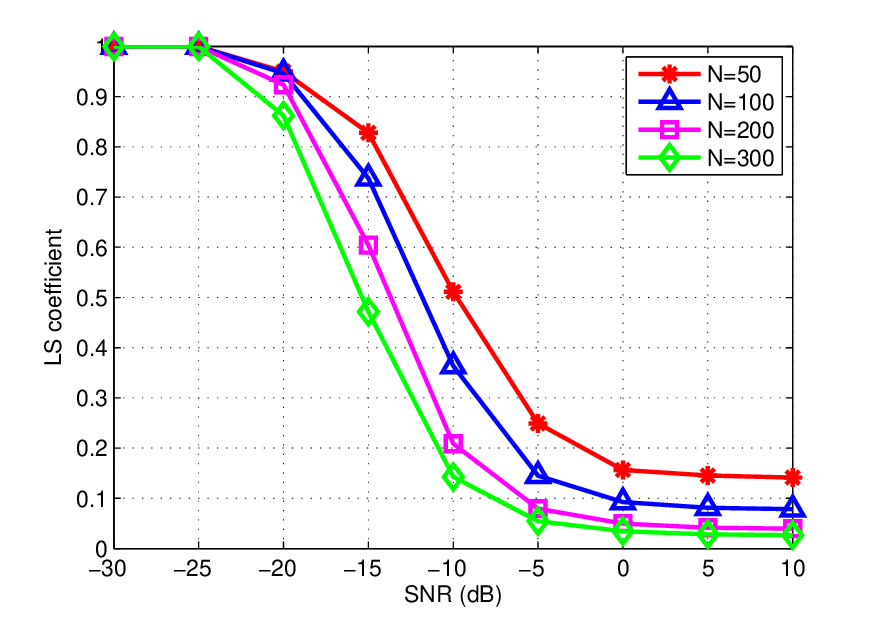} 
			\caption{LS coefficient estimate $\hat \rho$ versus SNR and snapshot number $N$, with $\bar M=40$, $K=2$ and $L=3$ .}\label{Fig.LS_coefficient}
		\end{figure}
		
		\begin{figure}[!t]
			\centering
			\includegraphics[width=3.5in]{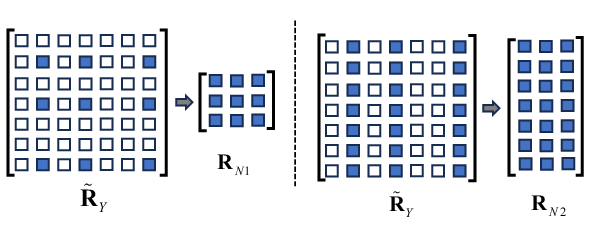} 
			\caption{Illustration of two special sub-matrices ${\bf{R}}_{N1}$ and ${\bf{R}}_{N2}$ randomly chosen from ${{{\bf{\tilde R}}}_Y}$.}\label{Fig.Napproximation}
		\end{figure}
		As shown in Fig.~\ref{Fig.LS_coefficient}, $\hat \rho$ is between 0 and 1, indicating that the constructed SCM ${{{\bf{\tilde R}}}_Y}$ can yield a good tradeoff between ${{{\bf{\hat R}}}_Y}$ and ${{\bf{R}}_T}$. In particular, for sufficiently low SNRs, $\hat \rho$ tends to 1 and ${{{\bf{\tilde R}}}_Y}={{\bf{R}}_T}$, and for high SNRs, $\hat \rho$ tends to 0 and ${{{\bf{\tilde R}}}_Y}\approx {{\bf{\hat R}}}_Y$. Such a characteristic guarantees the performance of the proposed solution in the whole SNR regions, as demonstrated later by both theoretical analysis and numerical simulations.

		With the enhanced SCM ${{{\bf{\tilde R}}}_Y}$ obtained, the number of total paths $KL$ can be estimated using either the LS-based two-step difference operation (SDO) \cite{ref24} or the adaptive diagonal loading (ADL) \cite{ref25} enumerators. A conventional approach would then perform eigenvalue decomposition (EVD) on ${{{\bf{\tilde R}}}_Y}$ to obtain the signal subspace required for MUSIC algorithm. However, when the virtual array size $\bar{M}$ is large,\textit{ direct EVD becomes computationally prohibitive.} To overcome this challenge while maintaining estimation accuracy, we adopt the Nystr\"{o}m approximation method \cite{ref26a1}, \cite{ref27a1} to efficiently approximate the ${\bar M}\times KL$ signal subspace matrix ${\bf U}_S$.
		
		The implementation of Nystr\"{o}m approximation involves extracting two specific sub-matrices from ${{{\bf{\tilde R}}}_Y}$, as illustrated in Fig.~\ref{Fig.Napproximation}:
		\begin{itemize}
			\item ${\bf{R}}_{N1} \in \mathbb{C}^{N_a \times N_a}$: The covariance matrix corresponding to a subset of $N_a$ randomly selected virtual antennas from the complete virtual ULA ($N_a \ll \bar{M}$).
			\item ${\bf{R}}_{N2} \in \mathbb{C}^{\bar{M} \times N_a}$: The cross-covariance matrix between the complete virtual ULA output and the output of those $N_a$ randomly selected antennas.
		\end{itemize}
		
		According to the principle of Nystr\"{o}m approximation, the following relationships hold
		\begin{equation}\label{a20}
			{\bf{R}}_{N1}\mathbf{u}_i=\gamma_i \mathbf{u}_i,\;{\bf{R}}_{N2}\mathbf{u}_i=\gamma_i\mathbf{u}_{ns,i},i=1,\dots ,N_a
		\end{equation}
		where $\left \{\gamma_i\right \}_{i=1}^{N_a}$ and $\left \{\mathbf{u}_i\right \}_{i=1}^{N_a} $ are eigenvalues and eigenvectors obtained by performing EVD on ${\bf{R}}_{N1}$, and $\left \{\mathbf{u}_{ns,i}\right \}_{i=1}^{N_a} $ are approximated values of principal eigenvector associated with ${{{\bf{\tilde R}}}_Y}$, which can be calculated by
		\begin{equation}\label{a21}
			\mathbf{u}_{ns,i}= \frac{\mathbf{R}_{N2}\mathbf{u}_i}{\gamma_i}.
		\end{equation}
		
		Subsequently, the approximate signal subspace matrix ${\bf{U}}_S$ with respect to ${{{\bf{\tilde R}}}_Y}$ can be constructed as
		\begin{equation}\label{a22}
			{\bf{U}}_S=\left [\mathbf{u}_{ns,1},\mathbf{u}_{ns,2},\dots ,  \mathbf{u}_{ns,KL}\right ].
		\end{equation}
		\begin{algorithm}[!t]
			\caption{TMRLS-MUSIC for DOA Estimation}\label{algorithm:al1}
			\LinesNumbered
			Move the designed FA array for ARS $G$ times within the $n$-th time block to construct ${\bf x}_n=[{\bf x}_{n,1}^T,\ldots,{\bf x}_{n,G}^T]$.\\
			Collect all the output ${\bf x}_n$ from  (\ref{5}) across $N$ consecutive time blocks to yield ${\bf{X}} = \left[ {{{\bf{x}}_1}, \ldots ,{{\bf{x}}_N}} \right]$.\\
			Rearrange the rows of ${\bf X}$ to form   ${\bf Y}$.\\
			Calculate    ${{\bf{\hat R}}_Y} = \frac{1}{N}{\bf{Y}}{{\bf{Y}}^H}$.\\
			Perform Toeplitz rectification to obtain ${{\bf{R}}_T}$ according to
			${{\bf{R}}_T} = \sum\nolimits_{m =  - \bar M + 1}^{\bar M - 1} {\frac{1}{{\bar M - \left| m \right|}}{\rm{Tr}}\left( {{{{\bf{\hat R}}}_Y}{{\bf{J}}^m}} \right){{\left( {{{\bf{J}}^T}} \right)}^m}} $.\\
			Calculate the LS coefficient ${\rho _a}$ according to  \eqref{18}.\\
			Obtain the enhanced SCM ${{{\bf{\tilde R}}}_Y} = \left( {1 - \hat \rho } \right){{{\bf{\hat R}}}_Y} + \hat \rho {{\bf{R}}_T}$.\\
			Apply the source enumeration algorithm to estimate the total number of paths $KL$.\\
			Extract two sub-matrices ${\bf{R}}_{N1}$ and ${\bf{R}}_{N2}$ from ${{{\bf{\tilde R}}}_Y}$.\\
			Perform EVD on ${\bf{R}}_{N1}$ to obtain $\left \{\gamma_i\right \}_{i=1}^{N_a}$ and $\left \{\mathbf{u}_i\right \}_{i=1}^{N_a}$.\\
			Calculate $\left \{\mathbf{u}_{ns,i}\right \}_{i=1}^{N_a} $ based on \eqref{a21} and then construct ${\bf{U}}_S$ based on \eqref{a22}.\\
			Obtain DOA estimates via the MUSIC spectral function:
			$f\left( \theta  \right) =\left [ {{{{\bf{b}}^H}\left( \theta  \right)(\mathbf{I} _{\bar M}-{{\bf{U}}_S}{\bf{U}}_S^H){\bf{b}}\left( \theta  \right)}} \right ]^{-1}$.
		\end{algorithm}
		
		Finally, the DOA estimates are obtained by finding the peaks of the following MUSIC spectral function
		\begin{equation}\label{20}
			f\left( \theta  \right) =\left [ {{{{\bf{b}}^H}\left( \theta  \right)(\mathbf{I} _{\bar M}-{{\bf{U}}_S}{\bf{U}}_S^H){\bf{b}}\left( \theta  \right)}} \right ]^{-1} ,\;\theta  \in ( - {90^ \circ },{90^ \circ }].
		\end{equation}
		
		The details of the proposed TMRLS-MUSIC algorithm are summarized in Algorithm \ref{algorithm:al1}.
		\begin{figure}[t]
			\centering
			\includegraphics[width=3.5in]{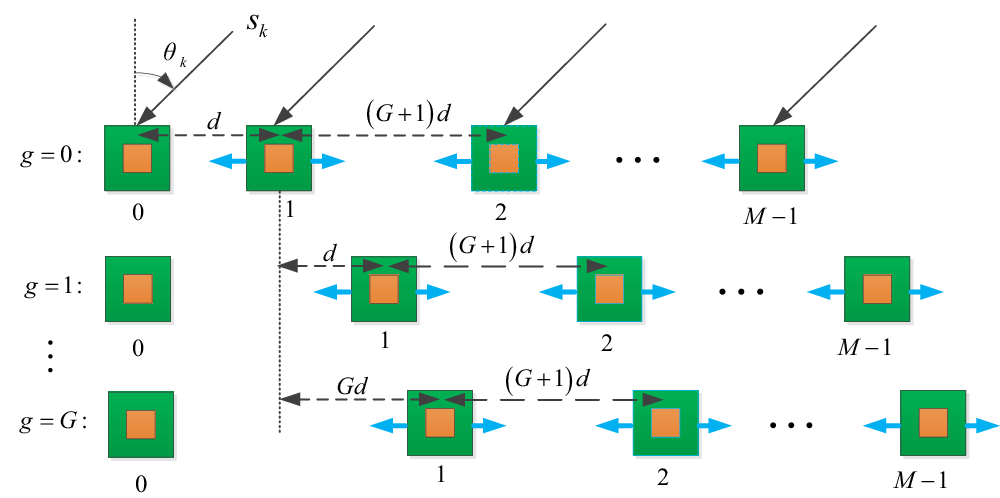}\\
			\caption{Illustration of FA array structure for NARS case.}\label{Fig.antennas_nonalign}
		\end{figure}
		\subsection{FA Array Design and DOA Estimation for NARS}\label{sec:NARS_model}

		For NARS case, the aligned signal model in \eqref{5} is invalid. To understand how NARS enables virtual array construction through covariance, we consider the fundamental principle behind this approach. In this scenario, while the signal content varies across positions (i.e., $s_{k,n,g_1} \neq s_{k,n,g_2}$ for $g_1 \neq g_2$), the covariance between signals received at different antenna positions preserves crucial phase information related to the position differences and signal directions.
		
		Specifically, for two antennas at positions $d_{m_1}$ and $d_{m_2}$ after movement $g$, the cross-covariance element is:
		\begin{align}\label{cov_principle}
			&{\bf{R}}_{xg}^{(m_1, m_2)} = \mathbb{E}\{x_{m_1,n,g} x_{m_2,n,g}^*\} \nonumber\\
			&= \sum_{k=1}^{K} \sum_{l=1}^{L} \sigma_{k,l}^2 e^{-j(d_{m_1,g} - d_{m_2,g})\varphi_{k,l}} + \sigma_e^2 \delta_{m_1,m_2},
		\end{align}
		where $\sigma_{k,l}^2 = \mathbb{E}\{|\alpha_{k,n,l}|^2\}$ is the power of the $l$-th path from the $k$-th source, and $\delta_{m_1,m_2}$ is the Kronecker delta function (equal to 1 when $m_1 = m_2$ and 0 otherwise).
		
		The key insight is that each covariance element ${\bf{R}}_{xg}^{(m_1, m_2)}$ contains the phase term $e^{-j(d_{m_1,g} - d_{m_2,g})\varphi_{k,l}}$, which depends only on the position difference $(d_{m_1,g} - d_{m_2,g})$ and the signal direction $\varphi_{k,l}$. This phase information is exactly what we need for DOA estimation, independent of the actual signal content. By collecting covariance elements corresponding to different position differences, we effectively construct a virtual array response that samples the spatial correlation function at various difference lags. The statistical averaging inherent in covariance computation removes the randomness of signal content while preserving the spatial structure, enabling DOA estimation even when the  signals cannot be directly aligned across different FA positions.
		
		This principle also explains why NARS requires a fundamentally different antenna configuration compared to ARS. If all antennas move together by the same amount (as in uniform translation), the position differences $(d_{m_1,g} - d_{m_2,g})$ remain constant across all movement states, resulting in zero phase variation in the covariance elements. Without phase variation, no spatial information can be extracted for DOA estimation. Therefore, NARS necessitates a fixed reference antenna that serves as a stable baseline, while other antennas move to create time-varying position differences. This design ensures that rich second-order difference lags are generated across different movement states, providing the necessary phase diversity to adequately sample the spatial correlation function and enable effective DOA estimation.
		
		Consequently, a different FA configuration, as depicted in Fig.~\ref{Fig.antennas_nonalign}, is employed to generate the necessary DoFs. This structure features one fixed reference antenna ($d_{1,g}=0, \forall g$) and $M-1$ FAs that follow the same movement pattern described in Section~\ref{sec:ARS_model}. 
		
		Since direct signal alignment is precluded in NARS, DOA estimation relies on the sub-covariance matrices corresponding to each movement state $g$. We introduce the concept of \textit{second-order difference lags}, which represent the relative position differences between any two antenna elements in the array. For example, if we have antennas at positions $d_1$, $d_2$, and $d_3$, the second-order difference lags would be $\\{d_2-d_1, d_3-d_1, d_3-d_2\\}$. Let ${\bf{X}}_g = \left[ {{\bf{x}}_{1,g}}, \ldots ,{{\bf{x}}_{N,g}} \right]$ be the data matrix for state $g$. Assuming equal-power transmitted signals that are independent of the path gains, the   sub-covariance matrix is
		\begin{equation}\label{21}
			{{\bf{R}}_{xg}} = \mathbb{E}\left\{ {{{\bf{X}}_g}{\bf{X}}_g^H} \right\} = {{\bf{A}}_g}{{\bf{R}}_{gS}}{\bf{A}}_g^H + \sigma _e^2{{\bf{I}}_M},
		\end{equation}
		where ${{\bf{R}}_{gS}}$ represents the covariance matrix of the transmitted signals at state $g$. Since   the transmitted signals have the same power and are independent of complex path gains, we have ${{\bf{R}}_{gS}}={{\bf{R}}_{S}}$ for all states of $g$. \textit{This means that while the instantaneous signal values may differ across positions, their statistical properties remain consistent, allowing us to extract the necessary phase information for DOA estimation through the covariance structure.}
		
		The key advantage of this approach is that it enables us to construct a virtual array response by carefully selecting and rearranging specific entries from these sub-covariance matrices. Each element of the virtual array response corresponds to one unique second-order difference lag, effectively sampling the spatial correlation function. This transformation from non-aligned signals to a virtual array response is what makes DOA estimation possible in the NARS scenario.

		\emph{Proposition 2:} The NARS FA configuration in Fig. \ref{Fig.antennas_nonalign} generates a set of unique second-order difference lags\footnote{A second-order difference lag $l_d$ corresponds to the difference ${g1} - l_{g2}$ between the integer positions $l_{g1}d$ and $l_{g2}d$ of any two elements in the array configuration for a given state $g$. These lags are implicitly available through the entries of the sub-covariance matrix ${\bf R}_{xg}$.} by combining the lags from all movement states $g=0, \dots, G$. Let ${\mathcal{L}}_D = \bigcup_{g=0}^G \{ (d_{m_1, g} - d_{m_2, g})/d \mid 1 \le m_1, m_2 \le M \}$ be the set of all achievable unique difference lags (normalized by $d$). Then, ${\mathcal{L}}_D$ contains $\tilde M = 2M_g + 1$ consecutive integer lags in the range $[-M_g, M_g]$, where $M_g = (M-1)(G+1)$.
		\begin{proof}
			For the movable $M-1$ units, the generated second-order differencing DoFs/lags are fixed with the range being the interval $\mathcal{S}_f=[ - (M - 2)(G + 1),(M - 2)(G + 1)]$ having a step of $G+1$, no matter how $g$ changes.
			
			The key to generating consecutive second-order difference lags lies in those generated between the fixed antenna and remaining $M-1$ movable units. After $g$ movements, the unique second-order differencing DoFs/lags generated between them are located in the interval $\mathcal{S}_g=[ - (M - 2)(G + 1) - g-1,(M - 2)(G + 1) + g+1]$ with a step size of $G+1$.
			Since $g$ changes continuously from 0 to $G$, it can be easily derived that ${\mathcal{S}_0} \cup {\mathcal{S}_1} \ldots  \cup {\mathcal{S}_G}$ yields $\tilde M = 2{M_g} + 1$ consecutive and unique second-order differencing DoFs/lags in the range $-{M_g} \le {l_p} \le {M_g}$. Since $\mathcal{L}_g={\mathcal{S}_g}\cup \mathcal{S}_f$, we   obtain that ${\mathcal{L}_P}$ contains $\tilde M = 2{M_g} + 1$ consecutive second-order differencing DoFs/lags in the range $-{M_g} \le {l_p} \le {M_g}$.
		\end{proof}
		
		\textbf{\textit{Proposition 2}} guarantees that the set of difference lags ${\mathcal{L}}_D$ covers the consecutive range $[-M_g, M_g]$. This allows us to construct a virtual array response vector ${\bf r}$ by carefully selecting and rearranging specific entries from the estimated sub-covariance matrices $\{{\bf{\hat R}}_{xg}\}_{g=0}^G$  where ${\bf{\hat R}}_{xg} = \frac{1}{N}{\bf X}_g {\bf X}_g^H$. Each element of ${\bf r}$ corresponds to one unique difference lag $l_d \in [-M_g, M_g]$. For instance, the element corresponding to lag $l_d = (d_{m_1, g} - d_{m_2, g})/d$ can be taken as the $(m_1, m_2)$ entry of ${\bf{\hat R}}_{xg}$. By vectorizing these selected entries in the order of lags from $-M_g$ to $M_g$, we obtain the vector ${\bf r} \in \mathbb{C}^{\tilde M \times 1}$. This vector effectively samples the spatial correlation function and, analogous to the ARS case, can be modeled as
		\begin{multline}\label{22}
			{\bf{r}} = [\underbrace {{\bf{\hat R}}_{xG}^{(1,M)}, \ldots ,{\bf{\hat R}}_{x0}^{(1,M)}}_{1 \times G},\underbrace {{\bf{\hat R}}_{xG}^{(1,M - 1)}, \ldots ,{\bf{\hat R}}_{x0}^{(1,M - 1)}}_{1 \times G}, \ldots , \\
			\;\underbrace {{\bf{\hat R}}_{xG}^{(1,2)}, \ldots ,{\bf{\hat R}}_{x0}^{(1,2)}}_{1 \times G},{\bf{\hat R}}_{x0}^{(1,1)},\underbrace {{\bf{\hat R}}_{x0}^{(2,1)}, \ldots ,{\bf{\hat R}}_{xG}^{(2,1)}}_{1 \times G}, \ldots ,\\
			\underbrace {{\bf{\hat R}}_{x0}^{(M,1)}, \ldots ,{\bf{\hat R}}_{xG}^{(M,1)}}_{1 \times G}{]^T} \approx {\bf{Cp}} + \sigma _e^2{{\bf{i}}_{{M_g} + 1}}\quad\quad\quad
		\end{multline}
		where ${\bf p} \in \mathbb{R}^{KL \times 1}$ contains the powers of the $KL$ signal paths (diagonal elements of ${\bf R}_S$), ${\bf i}_{\tilde M}$ is a vector containing 1 at the position corresponding to the zero lag ($l_d=0$) and zeros elsewhere (assuming noise only affects the zero lag correlation), and ${\bf C}$ is the $\tilde M \times KL$ virtual array manifold matrix for this difference array. The columns of ${\bf C}$ are steering vectors ${\bf c}(\theta_{k,l})$ corresponding to each path:
		\begin{equation}\label{23}
			{\bf{c}}(\theta) = {\left[ e^{-jM_g \varphi}, \dots, 1, \dots, e^{jM_g \varphi} \right]^T}
		\end{equation}
		with $\varphi=(2\pi/\lambda)\sin(\theta)$. This structure arises because the correlation corresponding to lag $l_d$ contains the phase term $e^{j l_d \varphi}$.
		
		We divide ${\bf{r}}$ into $M_g+1$ overlapping segments ${{\bf{r}}_1}, \ldots ,{{\bf{r}}_{{M_g} + 1}}$, then construct the Toeplitz SCM by the following operation
		\begin{multline}\label{24}
			\quad\quad{{\bf{R}}_C} = \left[ {{{\bf{\Pi }}_{{M_g} + 1}}{{\bf{r}}_1}, \ldots ,{{\bf{\Pi }}_{{M_g} + 1}}{{\bf{r}}_{{M_g} + 1}}} \right] \\
			= {\bf{\bar C}}{{\bf{R}}_S}{{\bf{\bar C}}^H} + \sigma _e^2{{\bf{I}}_{{M_g} + 1}},\quad\quad\quad\quad\quad\quad\quad\;\;
		\end{multline}
		where ${\bf{\bar C}}$ is a new array manifold matrix formed by the elements from the $(M_g+1)$-th row to $(2M_g+1)$-th row of ${\bf{C}}$.

		Applying the Nystr\"{o}m approximation again to obtain $(M_g+1)\times LK$ approximate signal subspace matrix ${\bf{\bar U}}_S$, the DOAs are finally estimated by the MUSIC technique (designated here as TMR-MUSIC) again.
		The main steps of TMR-MUSIC are
		summarized in Algorithm \ref{al2}.
		\begin{algorithm}[!t]
			\caption{TMR-MUSIC for DOA Estimation}\label{al2}
			\LinesNumbered
			Move the designed FA array for NARS $G$ times within the $n$-th time block to construct ${\bf x}_n=[{\bf x}_{n,1}^T,\ldots,{\bf x}_{n,G}^T]$.\\
			Collect and rearrange all the output ${\bf x}_n$, $n=1,\ldots,N$, to construct ${{\bf{X}}_g} = \left[ {{{\bf{x}}_{1,g}}, \ldots ,{{\bf{x}}_{N,g}}} \right]$.\\
			Calculate a series of sub-covariance matrices based on ${{{\bf{\hat R}}}_{xg}} = \frac{1}{N}{{\bf{X}}_g}{\bf{X}}_g^H,g = 0, \ldots ,G$.\\
			Rearrange the elements in series $\{{\bf{\hat R}}_{xg}\}_{g=0}^G$ to yield ${\bf{r}}$ according to \eqref{22}.\\
			Divide ${\bf{r}}$ into $M_g+1$ overlap segments, and construct the Toeplitz SCM ${{\bf{R}}_C} = \left[ {{{\bf{\Pi }}_{{M_g} + 1}}{{\bf{r}}_1}, \ldots ,{{\bf{\Pi }}_{{M_g} + 1}}{{\bf{r}}_{{M_g} + 1}}} \right]$.\\
			Apply the source enumeration algorithm to obtain the total number of paths $KL$.\\
			Extract two sub-matrices ${\bf{R}}_{N1}$ and ${\bf{R}}_{N2}$ from ${\bf{\bar C}}$.\\
			Perform EVD on ${\bf{R}}_{N1}$ to obtain $\left \{\gamma_i\right \}_{i=1}^{N_a}$ and $\left \{\mathbf{u}_i\right \}_{i=1}^{N_a}$.\\
			Calculate $\left \{\mathbf{u}_{ns,i}\right \}_{i=1}^{N_a} $ and subsequently construct ${\bf{\bar U}}_S$.\\
			Obtain DOA estimates via the MUSIC spectral function:
			$f\left( \theta  \right) =\left [ {{{{\bf{\bar c}}^H}\left( \theta  \right)(\mathbf{I} _{M_g+1}-{{\bf{\bar U}}_S}{\bf{\bar U}}_S^H){\bf{\bar c}}\left( \theta  \right)}} \right ]^{-1}$.
		\end{algorithm}
		\section{Performance Analysis}\label{performance}
		In this section, we analyze the performance of the proposed methods from four aspects: estimation accuracy, maximum number of estimable U/Ts, computational complexity, and capacity to localize U/Ts in different transmission scenarios.
		
		\subsection{Estimation Accuracy}
		The   TMRLS-MUSIC algorithm achieves significant improvement in estimation accuracy through an adaptive covariance matrix estimation approach. As illustrated in Fig.~\ref{Fig.LS_coefficient}, the LS coefficient $\hat{\rho}$ dynamically balances between the original  SCM ${{\bf{\hat R}}}_Y$ and the Toeplitz rectified matrix ${{\bf{R}}_T}$. In low SNR regions, $\hat{\rho}$ approaches 1, making ${{{\bf{\tilde R}}}_Y}$ equal to ${{\bf{R}}_T}$, which effectively overcomes the ``performance breakdown" phenomenon by enforcing the Toeplitz structure. Conversely,at high SNRs, $\hat{\rho}$ approaches 0, making ${{{\bf{\tilde R}}}_Y}$ approximately equal to ${{\bf{\hat R}}}_Y$, thereby avoiding the ``saturation phenomenon" by preserving the original signal characteristics.
		
		The estimation accuracy is further enhanced by the proposed FA array structures, which significantly increase the number of virtual sensors. For ARS scenarios, the number of virtual sensors increases from $M$ to $\bar M=MG + M$, while for NARS scenarios, it increases from $M$ to $M_g+1=(M-1)(G+1)+1$. This expansion in virtual sensors directly improves the estimation accuracy, as demonstrated by the Cram\'{e}r-Rao bound (CRB) analysis in \cite{ref29}. The enhanced spatial resolution enables better discrimination between closely spaced sources and more accurate DOA estimation.

		\subsection{Maximum Number of Estimable U/Ts}
		The proposed methods possess remarkable capability in estimating a large number of U/Ts through   effective utilization of FA mobility. According to the subspace theory, the maximum number of estimable paths is determined by the dimension of the enhanced SCMs, as at least one eigenvector must be reserved for spanning the noise subspace. For the proposed TMRLS-MUSIC algorithm, the enhanced SCM ${{{\bf{\tilde R}}}_Y}$ has dimensions of $\bar M \times \bar M$, enabling the estimation of up to $\bar M -1=MG+M-1$ paths. Similarly, for the TMR-MUSIC algorithm, the enhanced SCM ${{\bf{R}}_C}$ with dimensions $(M_g+1) \times (M_g+1)$ allows for estimating $M_g=(M - 1)(G + 1)$ paths. This capability for underdetermined DOA estimation is particularly valuable for discrete multipath channels, where the number of paths may exceed the number of physical antennas.
		
		\begin{figure*}[t]
			\centering
			\subfigtopskip=2pt
			\subfigbottomskip=-2pt
			\subfigcapskip=-5pt
			\subfigure[]{\includegraphics[width=3.3in]{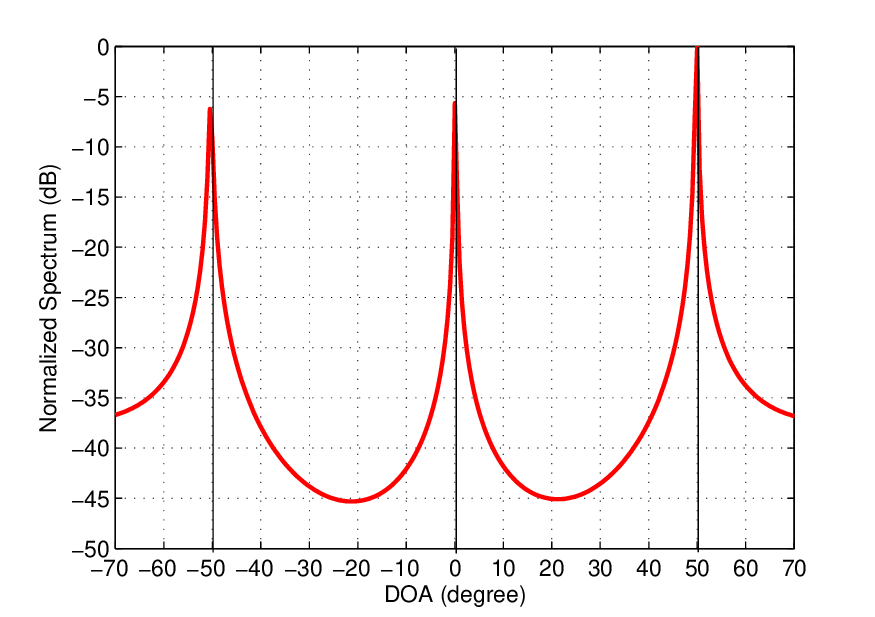}\label{fig1.a}}\hspace{-2mm}
			\subfigure[]{\includegraphics[width=3.3in]{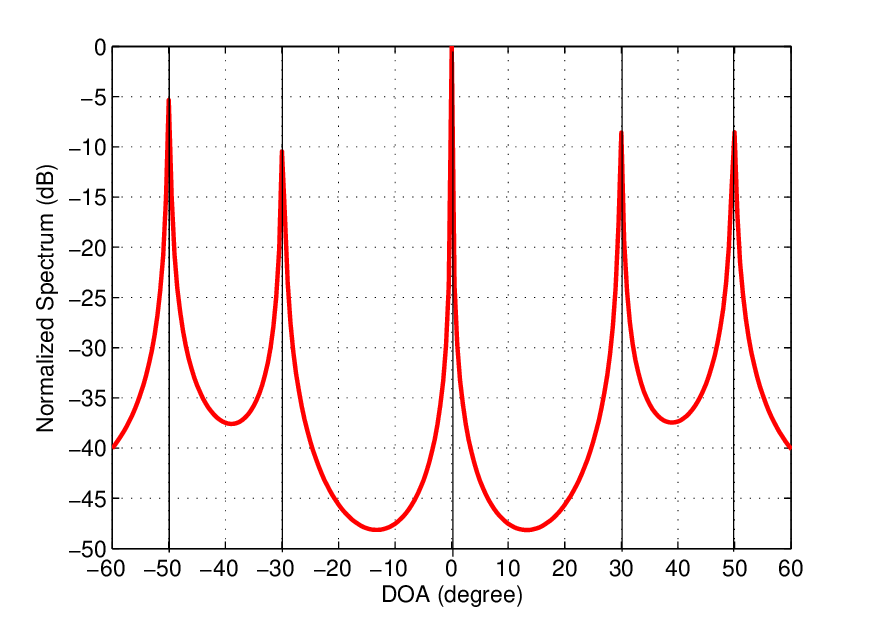}\label{fig1.b}}\hspace{-2mm}
			
			\subfigure[]{\includegraphics[width=3.3in]{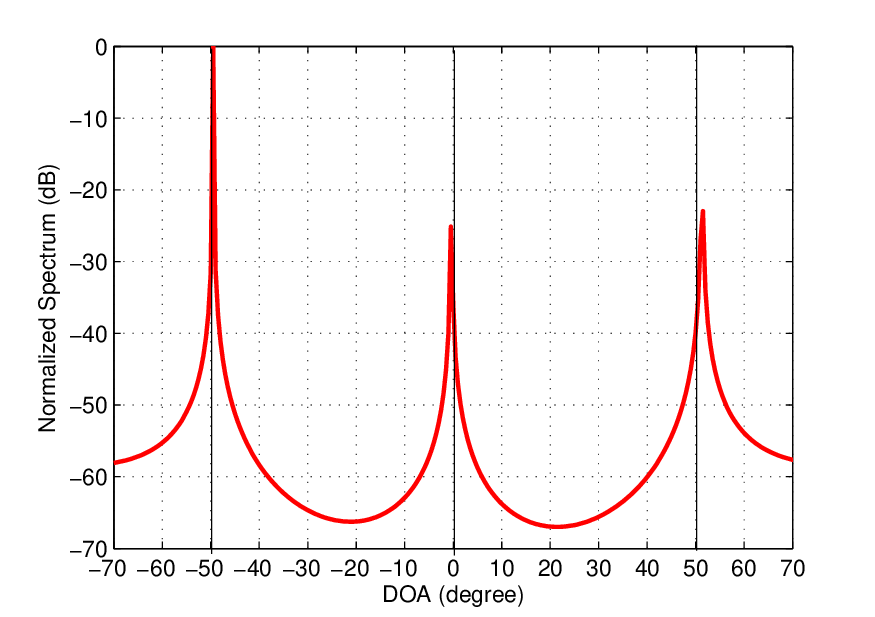}\label{fig1.d}}\hspace{-2mm}
			\subfigure[]{\includegraphics[width=3.3in]{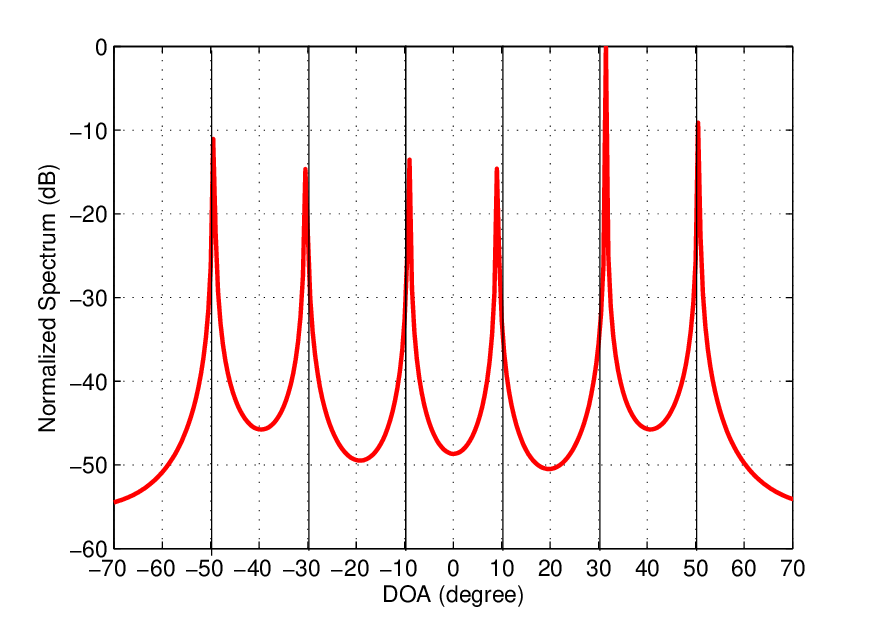}\label{fig1.f}}\hspace{-2mm}
			\caption{Spatial spectra of the proposed solutions under different configurations: (a) and (b) are for ANR; (c) and (d) are for NARS.}
			\label{fig:spectra1}
		\end{figure*}
		
		The maximum number of estimable U/Ts can be further quantified based on the system parameters. Given the number of physical antennas $M$, the number of movements $G$, and the number of paths per U/T $L$, the TMRLS-MUSIC algorithm can estimate up to $\left \lceil {(\bar M -1)}/{L}-1 \right \rceil$ U/Ts, while the TMR-MUSIC algorithm can handle $\left \lceil {M_g}/{L}-1 \right \rceil$ U/Ts. This enhanced capacity is achieved through the strategic movement of FA elements, where even a single FA in ARS scenarios or two FAs in NARS scenarios can estimate multiple DOAs by increasing the number of movements $G$. This remarkable feature effectively demonstrates the significant advantages that FA technology brings to the field of DOA estimation, particularly in scenarios where conventional fixed-position antenna arrays would be limited by their physical constraints.

		\subsection{Computational Complexity}
		In terms of computational complexity, we focus on the key component, including the multiplications required for SCM construction, LS coefficient calculation, EVD, and the MUSIC spectral search. For the TMRLS-MUSIC, the construction of the ${\bar M}\times {\bar M}$  SCM ${{{\bf{\tilde R}}}_Y}$, EVD on ${\bf{R}}_{N1}$ and calculation of LS coefficient require $\mathcal{O}(\bar MN^2+N_a^3+\bar M^3)$. In addition, the one-dimensional (1-D) MUSIC spectral search has the complexity of  $\mathcal{O}(\bar M^2KL+\bar M^2\bar G)$. Therefore, the proposed TMRLS-MUSIC requires $\mathcal{O}(\bar MN^2+N_a^3+\bar M^3+\bar M^2KL+\bar M^2\bar G)$, where $\bar G$ represents the number of searching grids. For the TMR-MUSIC, it calculates $G$ $M\times M$  sub-SCMs, and implements EVD on the $N_a\times N_a$ matrix ${\bf{R}}_{N1}$. Meanwhile, it also requires 1-D MUSIC spectral search once. Hence, the required total multiplication number is in order of $\mathcal{O}(GM^3+N_a^3+(M_g+1)^2KL+(M_g+1)^2\bar G)$. In comparison with the conventional MUSIC algorithm, the proposed two solutions are computationally more efficient, as the EVD of corresponding matrix is reduced from ${\bar M}\times {\bar M}$ and $(M_g+1)\times (M_g+1)$ dimensions are to ${N_a}\times {N_a}$, while the computations in other major parts remain the same.
		\begin{figure*}[t]
			\centering
			\subfigtopskip=0pt
			\subfigbottomskip=-2pt
			\subfigcapskip=-5pt
			\subfigure[]{\includegraphics[width=3.3in]{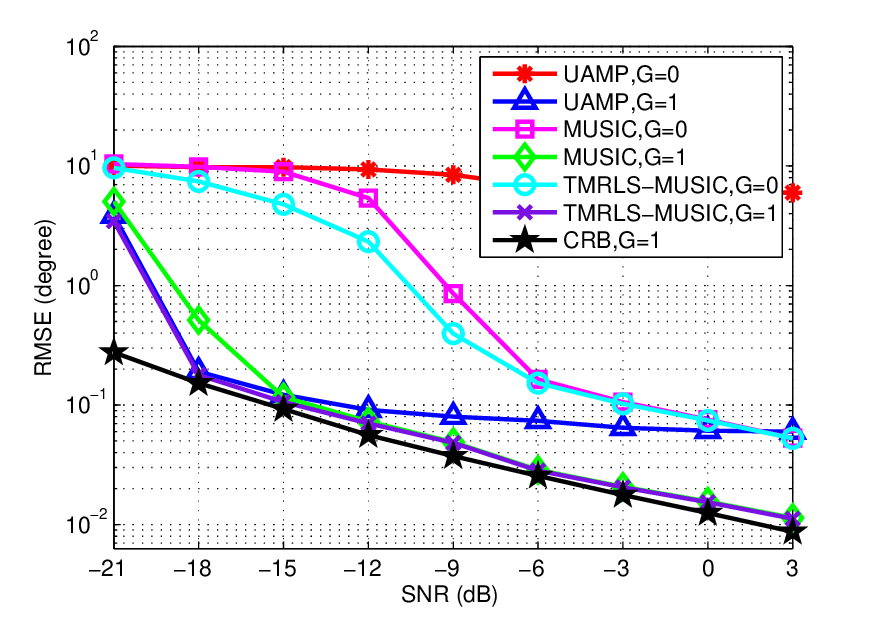}\label{fig2a}}\hspace{-2mm}
			\subfigure[]{\includegraphics[width=3.3in]{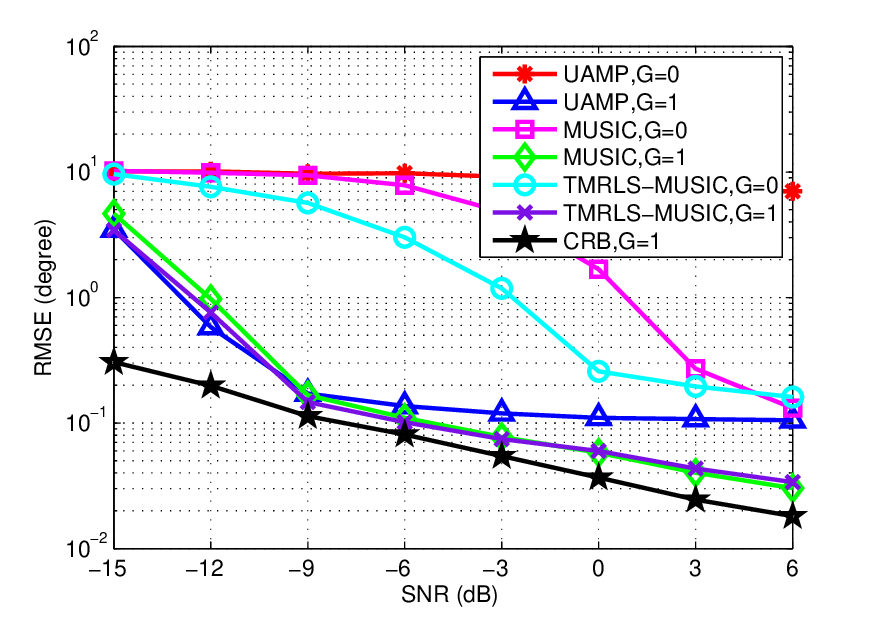}\label{fig2b}}\hspace{-2mm}
			\caption{RMSE of DOA estimations for ARS under various SNRs: (a) $M=20, N=200$; (b) $M=20, N=40$.}
			\label{fig:rmse_ars_snr}
		\end{figure*}
		\begin{figure*}[t]
			\centering
			\subfigtopskip=0pt
			\subfigbottomskip=-2pt
			\subfigcapskip=-5pt
			\subfigure[]{\includegraphics[width=3.3in]{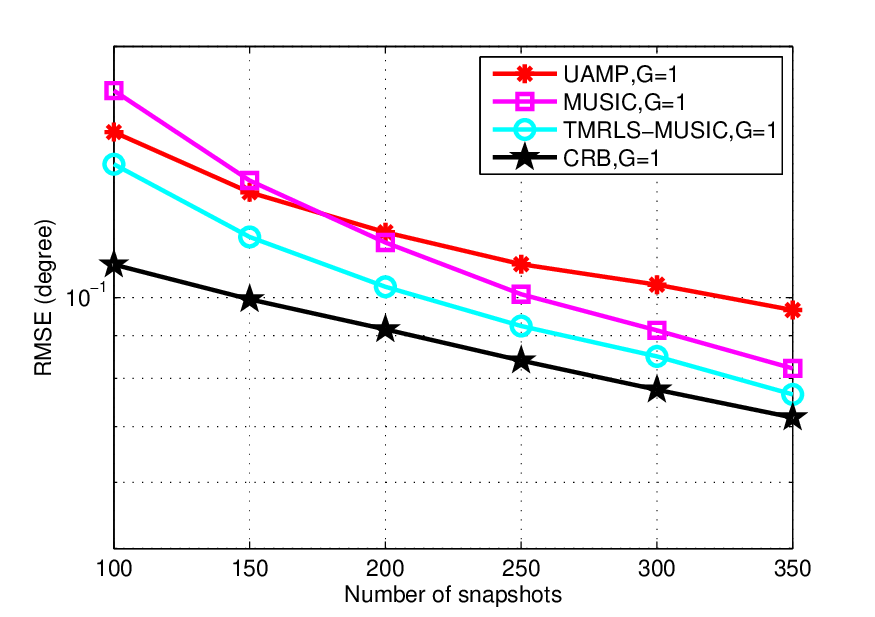}\label{fig3a}}\hspace{-2mm}
			\subfigure[]{\includegraphics[width=3.3in]{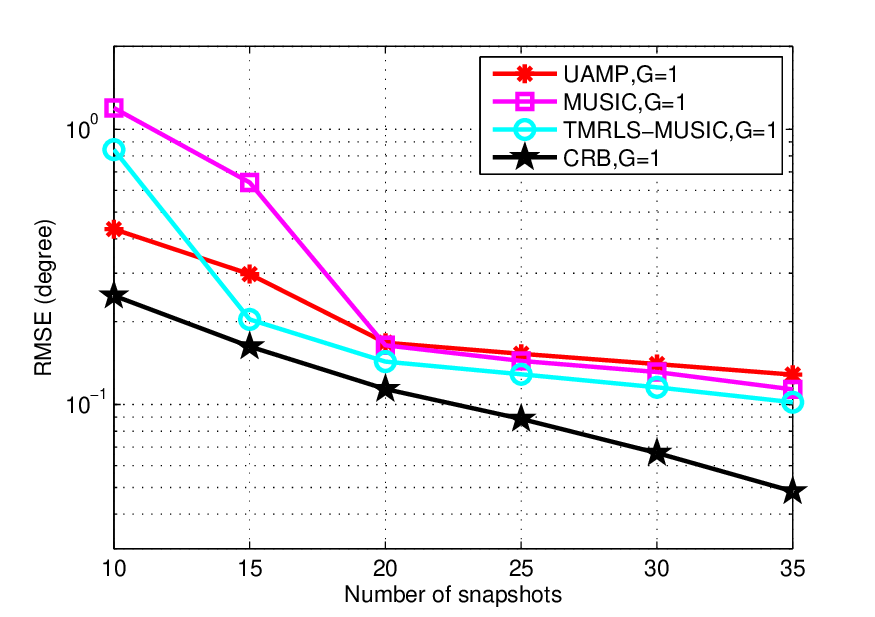}\label{fig4b}}\hspace{-2mm}
			\caption{RMSE of DOA estimations for ARS under different number of snapshots: (a) $M=20$, SNR=-15 dB; (b) $M=20$, SNR=-9 dB.}
			\label{fig:rmse_ars_snapshot}
		\end{figure*}

		\section{Simulation Results} \label{Simulation}
		This section evaluates the performance of the proposed FA-enabled DOA estimation methods: TMRLS-MUSIC for ARS and TMR-MUSIC for NARS. We compare them against conventional FPA-based MUSIC \cite{ref4} and  UAMP  algorithm \cite{ref7} adapted for FPAs. The  CRB obtained by utilizing SCM with extended DoFs is also selected as a benchmark. The root mean square error (RMSE) of DOA estimates based on  500 independent Monte-Carlo trials, is used to investigate the performance of the proposed solution corresponding to the extended virtual array aperture, which is defined as
		\begin{equation}\label{25}
			\mathrm{RMSE}   =\sqrt{\frac{1}{500KL}  {\textstyle \sum_{i=1}^{500}} {\textstyle \sum_{k=1}^{KL}( \hat \theta_{k,i} -\theta_k )^2 }  },
		\end{equation}
		where $\hat \theta_{k,i}$ is the estimate of $\theta_k$ for the $i$-th Monte-Carlo trial. 
		\begin{figure*}[t]
			\centering
			\subfigtopskip=0pt
			\subfigbottomskip=-2pt
			\subfigcapskip=-5pt
			\subfigure[]{\includegraphics[width=3.3in]{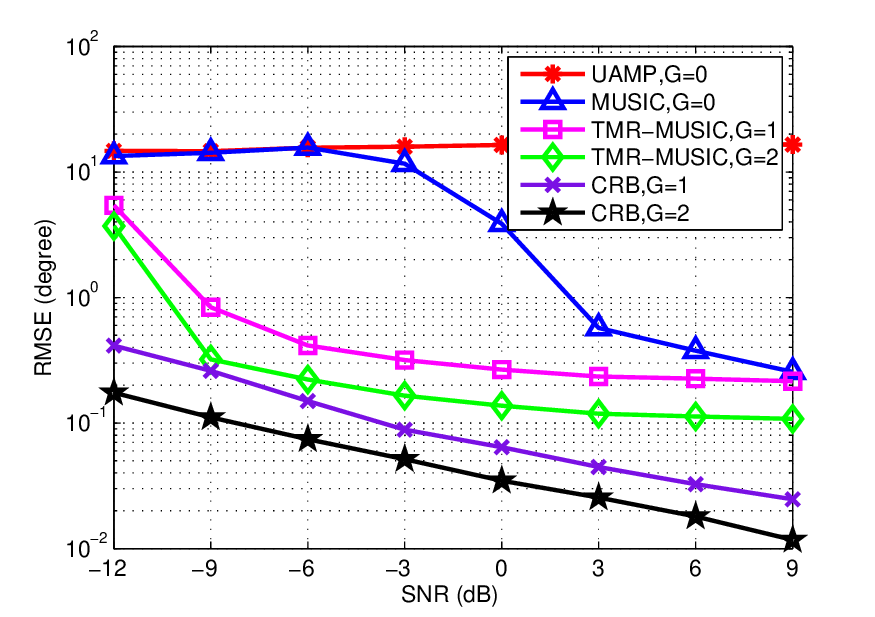}\label{fig5.a}}\hspace{-2mm}
			\subfigure[]{\includegraphics[width=3.3in]{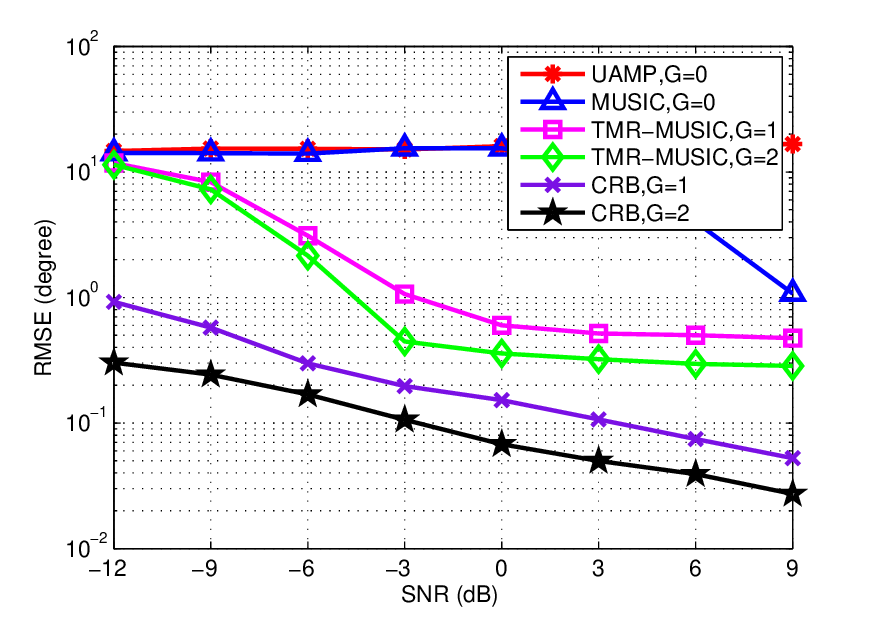}\label{fig5.b}}\hspace{-2mm}
			\caption{RMSE of DOA estimations for NARS under various SNRs: (a) $M=7, N=200$; (b) $M=7, N=40$.}
			\label{fig:rmse_nars}
		\end{figure*}
		\begin{figure*}[t]
			\centering
			\subfigtopskip=0pt
			\subfigbottomskip=-2pt
			\subfigcapskip=-5pt
			\subfigure[]{\includegraphics[width=3.3in]{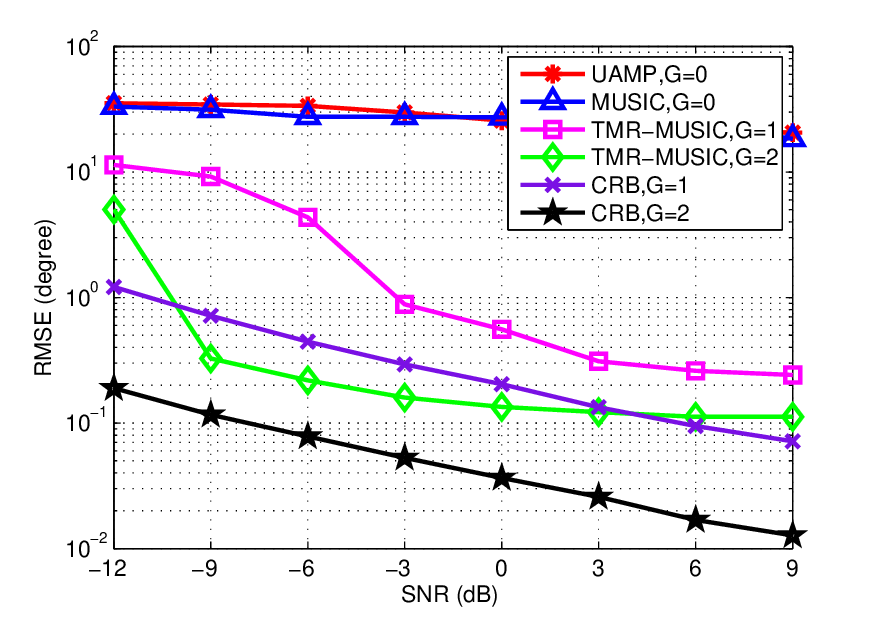}\label{fig5.a}}\hspace{-2mm}
			\subfigure[]{\includegraphics[width=3.3in]{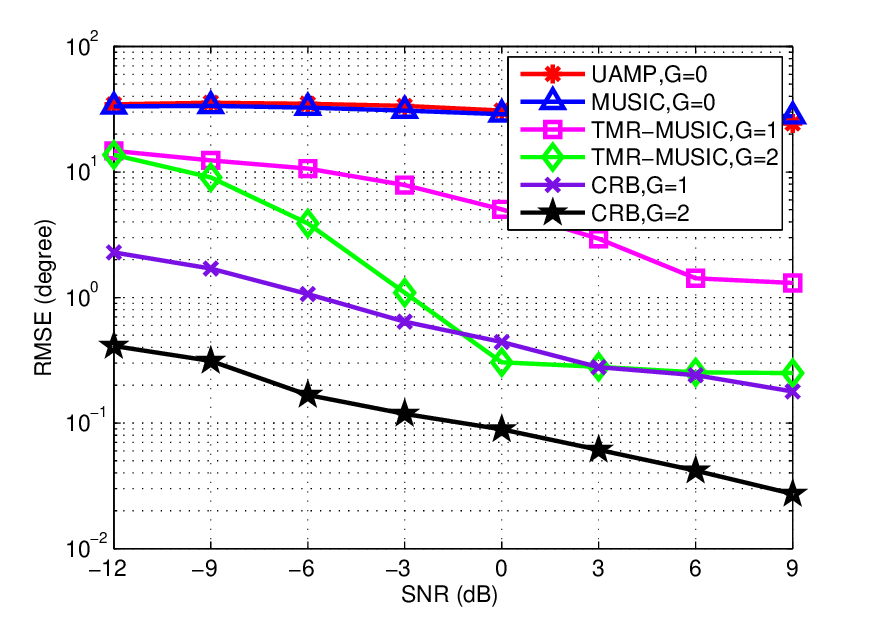}\label{fig5.b}}\hspace{-2mm}
			\caption{RMSE of DOA estimates versus SNR for ARS in closely-spaced scenarios: (a) $M=7, N=200$; (b) $M=7, N=40$.}
			\label{fig:rmse_nars_dense}
		\end{figure*}
		\begin{figure*}[h]
			\centering
			\subfigtopskip=2pt
			\subfigbottomskip=-2pt
			\subfigcapskip=-5pt
			\subfigure[]{\includegraphics[width=3.3in]{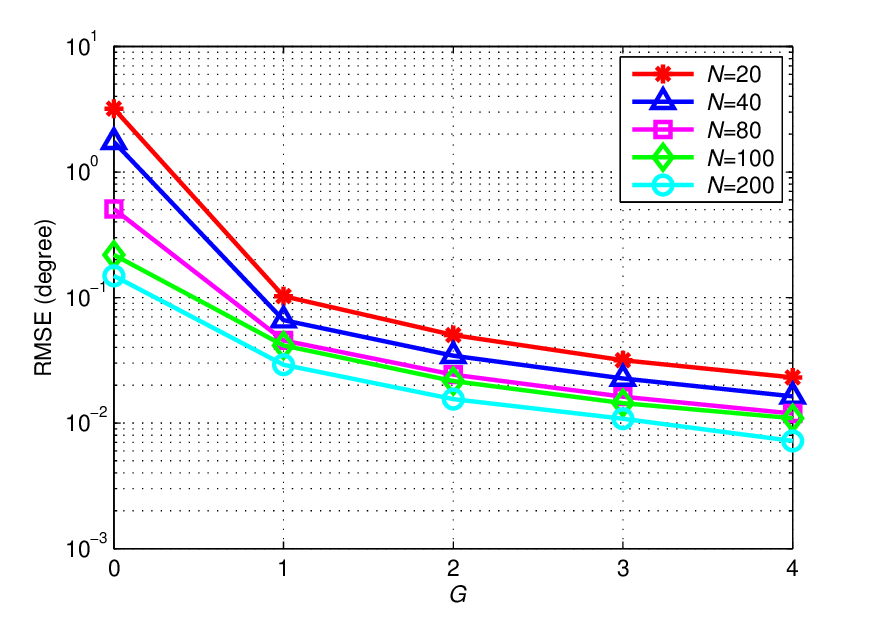}\label{fig3.a}}\hspace{-2mm}
			\subfigure[]{\includegraphics[width=3.3in]{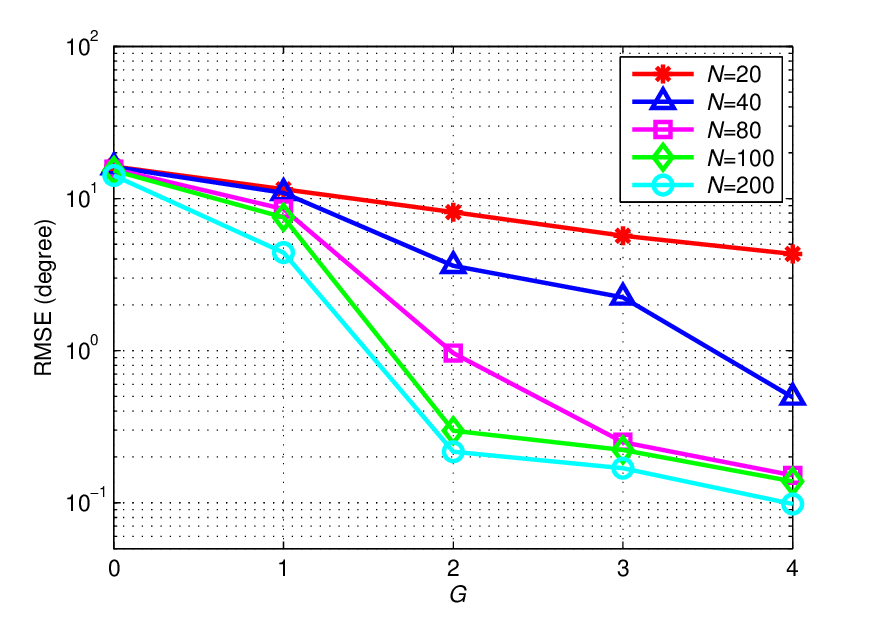}\label{fig3.b}}\hspace{-2mm}
			\caption{RMSE of DOA estimations under various $G$ and $N$: (a) is for ARS; (b) is for NARS.}
			\label{fig:spectra}
		\end{figure*}
		\subsection{Spatial Spectrum}
		First, we demonstrate the capability of the proposed methods for DOA estimation via their spatial spectra in Fig.~\ref{fig:spectra1}. The SNR is set to 10 dB and $N=200$ snapshots are used. Fig.~\ref{fig:spectra1}(a) plots the TMRLS-MUSIC spectrum for ARS with $M=1$ and $ G=3$, indicating that our approach exploit single FA to achieve estimation of multiple DOAs; Fig.~\ref{fig:spectra1}(b) further shows the TMRLS-MUSIC spectrum for ARS with $M=2$ and $G=2$, successfully resolving $KL=5$ paths located at $\{-50^\circ, -30^\circ, 0^\circ, 30^\circ, 50^\circ\}$, with a virtual array of size $\bar M = M(G+1) = 6$. Fig.~\ref{fig:spectra1}(c) plots the TMR-MUSIC spectrum for NARS with $M=2$ and $G=2$, from which we can see its  capacity to estimate multiple DOAs with only two FAs. Fig.~\ref{fig:spectra1}(d) displays the TMR-MUSIC spectrum for NARS with $M=3$ and $G=2$, resolving $KL=6$ paths at $\{-50^\circ, -30^\circ, -10^\circ, 10^\circ, 30^\circ, 50^\circ\}$. The number of consecutive difference lags is $\tilde M = 2(M-1)(G+1)+1 = 13$. Both scenarios reveal successful estimation even when the number of paths $KL$ exceeds the number of physical antennas $M$, validating the ability to estimate up to $\bar M-1$ and $M_g = (M-1)(G+1)$ sources for ARS and NARS, respectively,  in accordance with our analysis.
		\subsection{DOA Estimation Accuracy for ARS}
		We examine the RMSE performance versus SNR for the ARS scenario in Fig.~\ref{fig:rmse_ars_snr}. We consider $K=2$ U/Ts, each with $L=3$ paths, resulting in $KL=6$ sources with DOAs $\{-15.2^\circ, -10.5^\circ, -5.3^\circ; 4.1^\circ, 10.3^\circ, 15.4^\circ\}$  while   $M=20$, and $N_a=\bar M/2$  are assigned.  We compare the cases without movement ($G=0$) and with one movement ( i.e., $G=1$). Fig.~\ref{fig:rmse_ars_snr}(a) considers $N=200$ snapshots, while Fig.~\ref{fig:rmse_ars_snr}(b) considers a smaller sample size $N=40$. As expected, performance improves with increasing SNR and movement ($G=1$ outperforms $G=0$). Notably, the proposed TMRLS-MUSIC consistently achieves the lowest RMSE, with performance closely approaching the  root CRB, and significantly outperforms FPA-MUSIC and UAMP across all   SNRs, especially in the small sample regime ($N=40$).
		
		Next, we further assess the RMSE   under different number of snapshots $N$ for ARS scenario  in Fig.~\ref{fig:rmse_ars_snapshot}. The configuration of source parameters and the number of physical sensors are the same as in Fig.~\ref{fig:rmse_ars_snr}. In Fig.~\ref{fig:rmse_ars_snapshot}(a), SNR $= -15$ dB, $G=1$ and the number of snapshots $N$ varies from 100 to 350, while in Fig.~\ref{fig:rmse_ars_snapshot}(b), SNR $= -9$ dB, $G=1$ and $N$ varies from 10 to 35. It is observed that  the RMSE of the  TMRLS-MUSIC decreases with the increase of $N$, and it yields satisfactory estimation accuracy under both large and small snapshot conditions, verifying its superiority.
		\subsection{DOA Estimation Accuracy for NARS}
		Fig.~\ref{fig:rmse_nars} presents the RMSE performance versus SNR for the NARS scenario. Here, $K=1$ U/T with $L=3$ paths ($KL=3$) is considered, with DOAs of $\{-15.3^\circ, -5.3^\circ, 6.1^\circ\}$. The number of physical sensors is set to $M=7$, and $N_a=M_g/2$. We evaluate the performance for $G=0, 1, 2$ movements. Since ARS-based methods are inapplicable, we compare the proposed TMR-MUSIC ($G=1, 2$) against FPA-MUSIC and UAMP (both corresponding to $G=0$). It is seen that the  TMR-MUSIC significantly outperforms the FPA-based methods. Furthermore, increasing the number of movements from $G=1$ to $G=2$ further increase the estimation accuracy,  demonstrating the benefit of the enhanced DoFs generated by FA mobility in the NARS case.
		
		Fig.~\ref{fig:rmse_nars_dense} further plots the RMSE   inthe presence of closely spaced sources. The simulation configurations are the same with Fig.~\ref{fig:rmse_nars}, except that the DOAs of three sources are now  $\{-11.3^\circ, -5.3^\circ, 1.3^\circ\}$. We observe that as the DOA interval becomes smaller, the FPA-based MUSIC and UAMP algorithms fail (with their RMSEs exceeding $20^\circ$), while the proposed method still performs well in DOA estimation, once again demonstrating the latter  superiority.

		\subsection{{DOA Estimation Accuracy versus Movement Number}}
		We further evaluate the impact of number of movements $G$ for both ARS and NARS. The simulation configurations are the same with those in Sections V-B and V-C, respectively, except that SNR is fixed at -6 dB, $G$ is varied from 0 to 4, and $N$ is varied from 20 to 200. The results are plotted in Fig. \ref{fig:spectra}, from which we can see that the estimation performance increases significantly with  $G$. More interestingly, under certain scenarios, adding one movement brings greater benefits than increasing the snapshot size tenfold (for instance, the RMSE with $G=1$ and $N=20$ is lower that that with $G=0$ and $N=200$), further demonstrating the huge potential advantages of FAs in   DOA estimation.
		\section{Conclusion}\label{conclusions}
		In this paper, we proposed two fluid antenna (FA)-enabled DOA estimators designed to operate under limited-mobility-time constraints. For the aligned received signals (ARS) scenario, we developed a uniform FA structure where all antenna elements are movable, forming a virtual uniform linear array (ULA) with consecutive spatial sampling points. This configuration maximizes the array aperture and enables the direct application of conventional DOA estimation algorithms. For the non-aligned received signals (NARS) scenario, we proposed a specialized structure consisting of one fixed reference antenna combined with movable elements, explicitly designed to generate rich second-order difference lags between antenna positions. Leveraging these carefully designed FA configurations, our methods effectively exploit consecutive degrees-of-freedom (DoFs) in both scenarios. To overcome the challenges of processing high-dimensional data with limited samples resulting from FA mobility and time-varying channels, we introduced the TMRLS-MUSIC algorithm for ARS, which adaptively balances structured and unstructured covariance estimations, and the TMR-MUSIC algorithm for NARS, aimed at extracting phase information directly from the covariance structure rather than the signals themselves. Both algorithms incorporate the Nystr\"{o} approximation to significantly reduce computational complexity while preserving estimation accuracy. Theoretical analysis and extensive simulation results verify that the proposed approaches successfully perform underdetermined DOA estimation, provide superior performance and reduced computational complexity compared to existing schemes, and maintain robustness under varying operational conditions. These characteristics render our FA-enabled solutions particularly advantageous for practical applications, where traditional fixed-position antenna (FPA) arrays would be severely limited by physical constraints.

	\end{document}